%% file: SMP-19-005_temp.tex
\pdfoutput=1
\documentclass[11pt,twoside,a4paper,cmspaper,final,collab]{cms-tdr}
\def\svnVersion{cbd0dd1}\def\svnDate{2026/05/08}\def\cmsCernNoTag{CERN-EP-2025-242}\def\cmsCernDate{\today}\def\cmsMessage{Published in the Journal of High Energy Physics as \href{http://dx.doi.org/10.1007/JHEP05(2026)020}{\doi{10.1007/JHEP05(2026)020}.}}
\begin{document}\cmsNoteHeader{SMP-19-005}

\newlength\cmsFigWidth\setlength\cmsFigWidth{0.4\textwidth}
\newlength\cmsTabSkip\setlength{\cmsTabSkip}{1.7ex}

\providecommand{\cmsTable}[1]{\resizebox{\textwidth}{!}{#1}}
\newcommand{\ajj}{{\PGg}\text{jj}\xspace}
\newcommand{\mjj}{\ensuremath{m_\mathrm{jj}}\xspace}
\newcommand{\etaj}{\eta_{\mathrm{jj}}}
\newcommand{\cEWK}{\ensuremath{\alpha_{\text{EW}}}}
\newcommand{\cS}{\ensuremath{\alpha^{2}_{\text{S}}}}
\newcommand{\cw}{\ensuremath{c_{\PW}}\xspace}
\newcommand{\chwb}{\ensuremath{c_{\PH\PW\HepParticle{B}{}{}}}\xspace}
\newcommand{\etajl}{\ensuremath{\eta_{\text{j}_1}}\xspace}
\newcommand{\etajs}{\ensuremath{\eta_{\text{j}_2}}\xspace}
\newcommand{\detajj}{\ensuremath{\Delta\eta_{\text{jj}}}\xspace}
\newcommand{\cg}{\ensuremath{C_{\PGg}}\xspace}
\newcommand{\sitit}{\ensuremath{\sigma_{\text{i}\eta\text{i}\eta}}\xspace}
\newcommand{\xsec}{\ensuremath{\sigma_{\text{EW}\,\ajj}=202 \pm 7 \stat ^{+35}_{-32}\syst\unit{fb}}\xspace}
\newcommand{\ewsig}{\ensuremath{\sigma_{\text{EW}\,\ajj}}\xspace}
\newcommand{\ptg}{\ensuremath{\pt^{\PGg}}\xspace}
\newcommand{\muf}{\ensuremath{\mu_{\text{F}}}\xspace}
\newcommand{\mur}{\ensuremath{\mu_{\text{R}}}\xspace}
\newcommand{\pveto}{\ensuremath{\pt^{\text{veto}}}\xspace}
\title{Measurements of electroweak production of a photon in association with two jets in proton-proton collisions at \texorpdfstring{$\sqrt{s}=13\TeV$}{sqrt(s) = 13 TeV}}

\date{\today}

\abstract{The first observation of electroweak production of a photon in association with two forward jets in proton-proton collisions is presented. The measurement uses data recorded by the CMS experiment at the LHC during 2016--2018 at a center-of-mass energy of 13\TeV, corresponding to an integrated luminosity of 138\fbinv. The analysis is performed in a region enriched in photon production via vector boson fusion, with a requirement on the transverse momentum of the photon to exceed 200\GeV. The cross section is measured to be $202^{+36}_{-32}\unit{fb}$, at a significance with respect to the null hypothesis that exceeds five standard deviations. This is in agreement with the standard model prediction of $177^{+13}_{-12}\unit{fb}$. Differential cross sections are measured as a function of various observables. Limits are set on dimension-6 effective field theory operators that contribute to the $\PW\PW\PGg$ interaction. The observed 95\% confidence intervals for the corresponding Warsaw basis Wilson coefficients $\cw$ and $\chwb$ are $[-0.11, 0.16]$ and $[-1.6, 1.5]$, respectively.}

\hypersetup{
pdfauthor={CMS Collaboration},
pdftitle={Measurement of the electroweak production of one photon and two jets in proton-proton collisions at sqrt(s) = 13 TeV},
pdfsubject={CMS},
pdfkeywords={CMS, electroweak, vector boson fusion}}

\maketitle

\section{Introduction}
\label{intro}
Vector boson fusion (VBF) in proton-proton (pp) collisions~\cite{Rauch:2016pai} is a process that can provide unique insights into electroweak (EW) physics. In a VBF process, two vector bosons are radiated from the incoming quarks in the protons, and they then fuse to produce a third boson. The fusion process thus provides direct sensitivity to the triboson interaction. The initial quarks are deflected slightly from the beam direction, and appear as forward jets in the CMS detector. This leads to the typical VBF signature with two jets in the final state that are widely separated in pseudorapidity $\eta$ and with a large invariant mass of the dijet system $\mjj$. These jets are hereafter referred to as ``tagging jets''. In VBF processes, no color exchange is expected between the two incoming quarks. This results in a low hadronic activity within the pseudorapidity gap between the two tagging jets~\cite{PhysRevD.47.101,Baldenegro:2022kvj}, implying that additional jets in this region are usually from the showering of partons. The VBF processes are therefore also important tools to validate parton shower models.

\begin{figure*}[!ht]
\centering
      \includegraphics[width=0.32\textwidth]{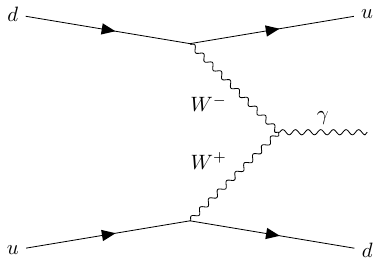}
        \caption{Representative Feynman diagram for  EW \ajj production with a photon produced via vector boson fusion.}   
\label{fig:ajj_feynmanvbf}
\end{figure*}

The large data sample of the CERN LHC at a center-of-mass energy of 13\TeV has provided the prime opportunity to better study the VBF production of a vector boson. The ATLAS and CMS Collaborations have measured the VBF production of the \PZ~\cite{CMS:2013fqx,ATLAS:2014sjq,CMS:2014sip,CMS:2017dmo,ATLAS:2017nei} and \PW~\cite{CMS:2016myt,ATLAS:2017luz,CMS:2019nep} bosons at different center-of-mass energies. Despite the larger cross section, the VBF production of a photon is more challenging to measure because of the overwhelming backgrounds that are irreducible and difficult to model. 

\begin{figure*}[!ht]
\centering
      \includegraphics[width=0.3\textwidth]{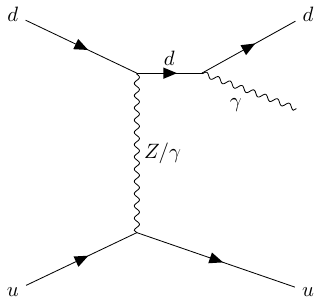}~~
      \includegraphics[width=0.28\textwidth]{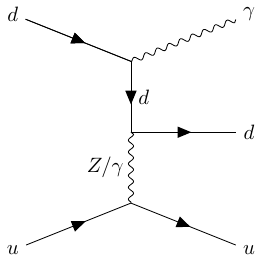}
	\caption{Representative Feynman diagrams for photons produced in FSR (left) and ISR (right).}   
\label{fig:ajj_feynman}
\end{figure*}

\begin{figure*}[!ht]
\centering
    \includegraphics[width=0.28\textwidth]{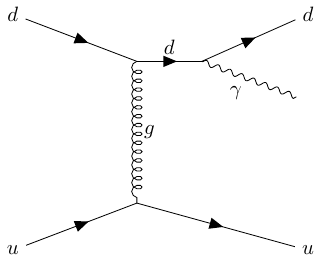}~~~~
    \includegraphics[width=0.28\textwidth]{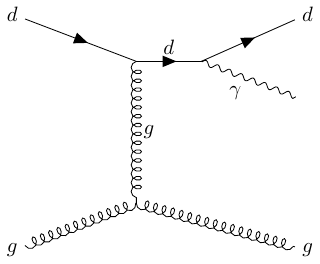} 
\caption{Representative Feynman diagrams for QCD-induced production of a photon and two jets.}
\label{fig:ajj_feynman2}
\end{figure*}

Figure~\ref{fig:ajj_feynmanvbf} shows a representative Feynman diagram for the EW \ajj vector-boson fusion process at order $\cEWK^2\alpha$, and Fig.~\ref{fig:ajj_feynman} shows EW \ajj productions via final-state radiation (FSR, left) or initial-state radiation (ISR, right). In Fig.~\ref{fig:ajj_feynman2}, diagrams of order $\cS\alpha$ are presented where the same final state is realized through dijet production induced by quantum chromodynamics (QCD) with a hard FSR photon. Such processes are among the irreducible QCD backgrounds with a production cross section that is about 30 times larger than EW \ajj production.
The ATLAS Collaboration has conducted detailed studies of photon production in association with two jets, focusing on the fragmentation and direct production regions~\cite{ATLAS:2019iaa}. To date, no dedicated measurement of purely EW \ajj production has been performed at the LHC.

In this paper, the first measurement of the EW \ajj production is presented. The measurement is performed in a VBF region and the photon is required to be central, $\abs{\eta}<1.44$, and have a transverse momentum greater than 200\GeV. It is based on $\Pp\Pp$ collisions at a center-of-mass energy of 13\TeV, recorded in 2016--2018 with the CMS detector at the LHC, corresponding to an integrated luminosity of 138\fbinv. The QCD-induced \ajj production is treated as an irreducible background. The signal interference with the QCD-induced processes, modeled at leading order (LO), is also treated as background. This study provides, in addition to the inclusive cross section, $\ewsig$, differential cross section measurements, unfolded to the particle level, as functions of pseudorapidity of the tagging jets \etajl and \etajs, the transverse momentum of the photon $\pt^\PGg$, the event centrality \cg~\cite{PhysRevD.92.012006}, the invariant mass of the two tagging jets \mjj, and the Zeppenfeld variable~\cite{Rainwater:1996ud}, which measures the photon’s pseudorapidity relative to the dijet system. The hadronic activity within the pseudorapidity gap between the two tagging jets is also studied to validate the simulation. Finally, limits are derived on new {\PW}{\PW}{\PGg} interactions described by the Warsaw basis~\cite{Grzadkowski:2010es} Wilson coefficients, \cw and \chwb, in the context of an effective field theory (EFT) extension to the standard model, SMEFT~\cite{Ellis:2021kzk}. These results are also tabulated in the HEPData record for this analysis~\cite{hepdata}.

The paper is organized as follows: a description of the CMS detector is provided in Section~\ref{cms_dec}. Simulations of signal and background processes are detailed in Section~\ref{samples}. The reconstruction and selection of events are described in Section~\ref{event_selection}. The estimation of the backgrounds is detailed in Section~\ref{bkg_estimation}. The signal-to-background separation is achieved via a multivariate discriminator described in Section~\ref{bdt}. A review of the systematic uncertainties is provided in Section~\ref{uncertainties}. Section~\ref{significance} presents the inclusive and differential measurements, and the EFT interpretation is the topic of Section~\ref{eft}. A summary is given in Section~\ref{summary}.

\section{The CMS detector}
\label{cms_dec}

The central feature of the CMS~\cite{Chatrchyan:2008zzk,Hayrapetyan_2024} apparatus is a superconducting solenoid of 6\unit{m} internal diameter, providing a magnetic field of 3.8\unit{T}. Within the solenoid volume are a silicon pixel and strip tracker, a lead tungstate crystal electromagnetic calorimeter (ECAL), and a brass and scintillator hadron calorimeter (HCAL), each composed of a barrel and two endcap sections. The ECAL provide a coverage in pseudorapidity $\abs{\eta} < 1.48 $ in the barrel region and $1.48 < \abs{\eta} < 3.0$ in two endcap regions. Forward calorimeters extend the $\eta$ coverage provided by the barrel and endcap detectors. Muons are detected in gas-ionization chambers embedded in the steel flux-return yoke outside the solenoid.

Events of interest are selected using a two-level trigger system~\cite{trigger_system,CMS:2024aqx}. The first level (L1), composed of custom hardware processors, uses information from the calorimeters and muon detectors to select events of interest with a maximum rate of 100\unit{kHz} within a latency of about 4\mus. The second level is a high-level trigger (HLT) processor, made up of a farm of processors running a version of the full event reconstruction software optimized for fast processing, and decreases the rate to a few \unit{kHz} before storage.

A more detailed description of the CMS detector, together with a definition of the coordinate system and kinematic variables, is reported in Ref.~\cite{Chatrchyan:2008zzk}.

\section{Signal and background simulation}
\label{samples}
{\tolerance=2500
The signal sample is generated using the Monte Carlo (MC) event generator \MGvATNLO version~2.6.5 (MG5)~\cite{MGatNLO} at next-to-leading-order (NLO) accuracy in perturbative QCD. The photon produced at the matrix-element level is required to be isolated in a cone of size 0.05~\cite{Frixione:1998jh}. The QCD-induced photon production with additional jets (QCD \ajj) is also modeled with MG5 at NLO. Samples are generated in bins of $\pt^\PGg$ to increase the number of events with a high-\pt photon. The interference between EW \ajj and QCD \ajj is modeled at LO with MG5. 
\par}
Other backgrounds arise from {\PW}{\PGg} and {\PZ}{\PGg} production with additional jets (V{\PGg}+jets) where the massive vector bosons decay hadronically, the production of diphoton plus jets ({\PGg}{\PGg}+jets) with a photon failing the selection requirements, and $\ttbar\PGg$ production with hadronically decaying top quarks. These processes are all generated with MG5 at NLO except $\PGg\PGg$+jets that is produced at LO with \textsc{Sherpa 2.2}~\cite{10.21468/SciPostPhys.7.3.034}. Parton showering and hadronization are modeled using \PYTHIA~8.240~\cite{Sjostrand:2014zea}, with the CP5 tune~\cite{Khachatryan:2015pea}, where the dipole recoil scheme~\cite{Jager:2020hkz} is employed for the EW \ajj signal. Parton diatribtion functions (PDFs) of NNPDF~3.1~\cite{NNPDF:2017mvq} are used for all samples. The detector response is simulated using $\GEANT4$~v10.4.3~\cite{geant4}. Simulated events are reweighted to have the same profile as in data for additional pp interactions in the same and neighboring bunch crossings (pileup).

\section{Event reconstruction and selection}
\label{event_selection}
Collision events are selected at the HLT according to the presence of a photon with $\pt>200\,(175)\GeV$ for the 2017--2018 (2016) data taking periods. The same requirements are imposed on simulated events and the trigger efficiencies are corrected as a function of \ptg to obtain the same trigger efficiency as in data. The trigger efficiencies in the data are obtained using event samples passing uncorrelated muon triggers, while containing a photon in the ECAL barrel with $\ptg > 200\GeV$. 
 
Events in data and simulation go through the same reconstruction and selection procedures. The particle-flow algorithm~\cite{Sirunyan:2017ulk} is employed to reconstruct and identify individual particles in the event, combining information from different CMS subdetectors. The primary vertex (PV) is taken to be the vertex corresponding to the hardest scattering in the event, which refers to the collision with the highest energy transfer during the event, identified using the tracking information alone, as described in Ref.~\cite{CMS-TDR-15-02}. 

Photons are identified as ECAL energy clusters not linked to the extrapolation of any charged particle trajectory from the silicon tracker. In the barrel section of the ECAL, an energy resolution of about 1\% is achieved for unconverted or late-converting photons in the tens of \GeV energy range. The energy resolution of the remaining barrel photons is about 1.3\% up to $\abs{\eta} = 1$, gradually increasing to about 2.5\% at $\abs{\eta} = 1.4$. 

Photons are selected in the barrel region, \ie, $\abs{\eta}<1.44$ with $\pt>200\GeV$. A set of identification criteria are imposed~\cite{CMS:2020uim,CMS:2018rym} on various photon properties, including, \eg, the spread in $\eta$ of the electromagnetic shower in the ECAL, \sitit, the ratio of deposited energy in the HCAL to ECAL, and the isolation, defined as the sum of transverse momenta of charged hadrons, $I_{\text{ch}}$, photons, $I_{\PGg}$, and neutral hadrons, $I_{\text{n}}$, within a cone of $\Delta R = \sqrt{\smash[b]{(\Delta\eta)^2+(\Delta\phi)^2}} = 0.3$ around the photon direction, excluding the photon itself~\cite{Sirunyan:2017ulk,CMS:2020uim,CMS:2018rym}. The ``tight'' and ``loose'' identification requirements correspond, respectively, to an average efficiency of 70\% and 90\%~\cite{Rembser_2019}. Photons must meet the tight identification criteria and leave no signal (seed) in the pixel detector. Corrections are applied to simulated events, as a function of photon \pt and $\abs{\eta}$, in order to achieve the same efficiency as in data for photon identification and the rejection of photons with seeds in the pixel detector. 

For each event, hadronic jets are clustered from the reconstructed particles using the infrared- and collinear-safe anti-\kt algorithm~\cite{Cacciari:2008gp, Cacciari:2011ma} with a distance parameter of 0.4. Jet momentum is determined as the vectorial sum of all particle momenta in the jet, and is found from simulation to be, on average, within 5 to 10\% of the true momentum over the whole \pt spectrum and detector acceptance. Particles from pileup can contribute additional tracks and calorimeteric energy deposits to the jet momentum. To mitigate this effect, charged particles identified to be originating from pileup vertices are discarded and an offset correction is applied to correct for remaining contributions~\cite{Sirunyan:2017ulk}. Jet energy corrections are derived from simulation to bring the measured response of jets to that of particle-level jets on average. In situ measurements of the momentum balance in dijet, $\text{photon} + \text{jet}$, $\PZ + \text{jet}$, and multijet events are used to account for any residual differences in the jet energy scale between data and simulation~\cite{CMS:2016lmd}. The jet energy resolution amounts typically to 20--30\% at 15\GeV, 15--20\% at 30\GeV, 10\% at 100\GeV, and 5\% at 1\TeV~\cite{CMS:2016lmd}. Additional selection criteria are applied~\cite{CMS-PAS-JME-16-003} to remove jets potentially dominated by anomalous contributions from various subdetector components or reconstruction failures. Jets are required to have $\pt>50\GeV$ and $\abs{\eta}<4.7$ and to be separated from one another with $\Delta R >0.4$.

Events are selected with at least one photon and at least two jets separated from the photon by more than 0.4 in $\Delta R$. The leading and subleading jets in \pt ($\text{j}_{1}$ and $\text{j}_{2}$, respectively), are considered as tagging jets and must be separated in pseudorapidity, $\abs{\detajj}>2.5$, while passing $\mjj > 500 \GeV$. Similar requirements are imposed on the EW \ajj signal at particle level to define the fiducial region.

\section{Background determination}
\label{bkg_estimation}
Backgrounds to the EW \ajj measurement can be divided into two main categories based on the photon origin. Processes containing prompt photons from the primary vertex are estimated using simulation. This category includes QCD \ajj, the interference between this process and the signal, the production of massive vector bosons and \ttbar in association with photons, and diphoton production. All processes are initially normalized to the theoretical prediction of their cross section at NLO, except for the diphoton production, which is calculated at LO.
 
The other category contains reducible backgrounds with nonprompt photons from hadron decays, or of purely instrumental origins. The contribution of nonprompt photons is estimated using an ABCD method~\cite{Fake}. Besides the signal region (A) described in Section~\ref{event_selection}, three orthogonal regions in data (B, C, and D) are introduced by modifying the photon identification criteria and/or the $\mjj$ requirement. Region B is similar to Region A except for the photon identification. It must pass the loose identification criteria while failing the condition on any of $I_{\text{ch}}$, $I_{\PGg}$, and $I_{\text{n}}$ variables. The threshold on $\sitit$ is also set to be less stringent that of the loose identification~\cite{CMS:2020uim}. These adjustments are designed to reject most prompt photons while increasing the acceptance for nonprompt candidates. Region D (C) is derived from region A (B) by modifying the $\mjj$ requirement to be less than 400\GeV. 

The correlation between the modified photon identification variables and $\mjj$ is checked in simulation and found to be negligible. Consequently, \textit{f}, the fraction of nonprompt photons passing the tight requirements, is expected to remain consistent between regions B and C. This allows the \textit{f} fraction, measured in regions C and D with $\mjj<400\GeV$, to be applied to region B for estimating the nonprompt photon contribution in signal region A.

Although small, the contribution of prompt photons must be subtracted from regions with modified photon identification criteria. This is achieved by fitting to the data the $\sitit$ distributions that differs between prompt and nonprompt photons. 
The $\sitit$ distribution for prompt photons is taken from simulation. Distributions from EW \ajj signal and QCD \ajj samples are alternatively used in the fit and the difference between the results is taken as a systematic uncertainty. The $\sitit$ distribution of nonprompt photons is extracted from a sideband in data. To define the sideband region, the tight requirement on photon \sitit is relaxed and the condition on $I_{\text{ch}}$ is inverted. To assess the robustness of the procedure, alternative sidebands are defined by varying the lower bound on $I_{\text{ch}}$. The corresponding $\sitit$ distributions are used in the fit and the variations in the results are taken as extra systematic uncertainties.
The \textit{f} fraction, and hence the nonprompt contribution in the signal region, is measured in bins of photon \pt. The closure of the method is verified by repeating the entire procedure using MC simulation. Residual nonclosures are considered as additional systematic uncertainties, alongside uncertainties arising from the limited size of the MC samples.

Table \ref{tab:prefit_yield} contains the estimation of the nonprompt photon background together with the expected event yields in the selected samples of prompt photons. The EW \ajj signal is divided by being inside or outside the signal phase space at generator level. Considering the total uncertainty, arising from both MC statistical limitations and systematic effects, the total number of simulated events is compatible with the observed data. 

\begin{table}[ht!]
\centering
\caption{Expected event yields and their uncertainties for signal and backgrounds, including also the estimation of the nonprompt photon contribution. The number of observed data events are also included for comparison.}

\newcolumntype{x}{D{,}{\,\pm\,}{12.6}}
\begin{tabular}{lx}
Process & \text{Events} \\
\hline
EW $\PW\PGg$jj (signal) & 9824,74 \\
EW $\PGg$jj (out of phase space) & 686,44 \\
QCD $\PGg$jj & 235689,8330 \\
EW-QCD $\PGg$jj (interference) & 428,31 \\
Nonprompt photon & 6052,493 \\
$\PGg \PGg$ & 1262,78 \\
$\PW\PGg\to\text{jj}\PGg$ & 1921,144 \\
$\PW\PGg\to\Pell\PGn\PGg$ & 713,56 \\
$\PZ\PGg\to\text{jj}\PGg$ & 729,54 \\

$\ttbar\PGg\text{ j}$ & 472,36 \\

Total expected & 257776,8347 \\[\cmsTabSkip]

Data & 248390,498 \\
\end{tabular}
      \label{tab:prefit_yield}
\end{table}

\section{Signal and background discrimination}
\label{bdt} 

In order to discriminate the EW \ajj signal from the dominant QCD \ajj background, a boosted decision tree (BDT) with gradient boosting~\cite{xgboost} is trained and optimized on simulated events passing the selection criteria detailed in Section~\ref{event_selection}. Inputs to the BDT are \etajl, \etajs, \detajj, \mjj, the transverse momentum of the leading jet $\pt{(\text{j}_1)}$, the angular separation between the photon and the subleading jet $\Delta R(\text{j}_2,\PGg)$, $C_{\PGg}$, the ratio between $\ptg$ and the scalar \pt sum of the tagging jets, and the Zeppenfeld variable defined as:
\begin{equation}
	\text{Zeppenfeld} = \lvert \eta_{\PGg} - \frac{\etajl + \etajs}{2} \rvert.
\end{equation}

\begin{figure}[ht!]
   \centering
      \includegraphics[width=0.49\textwidth]{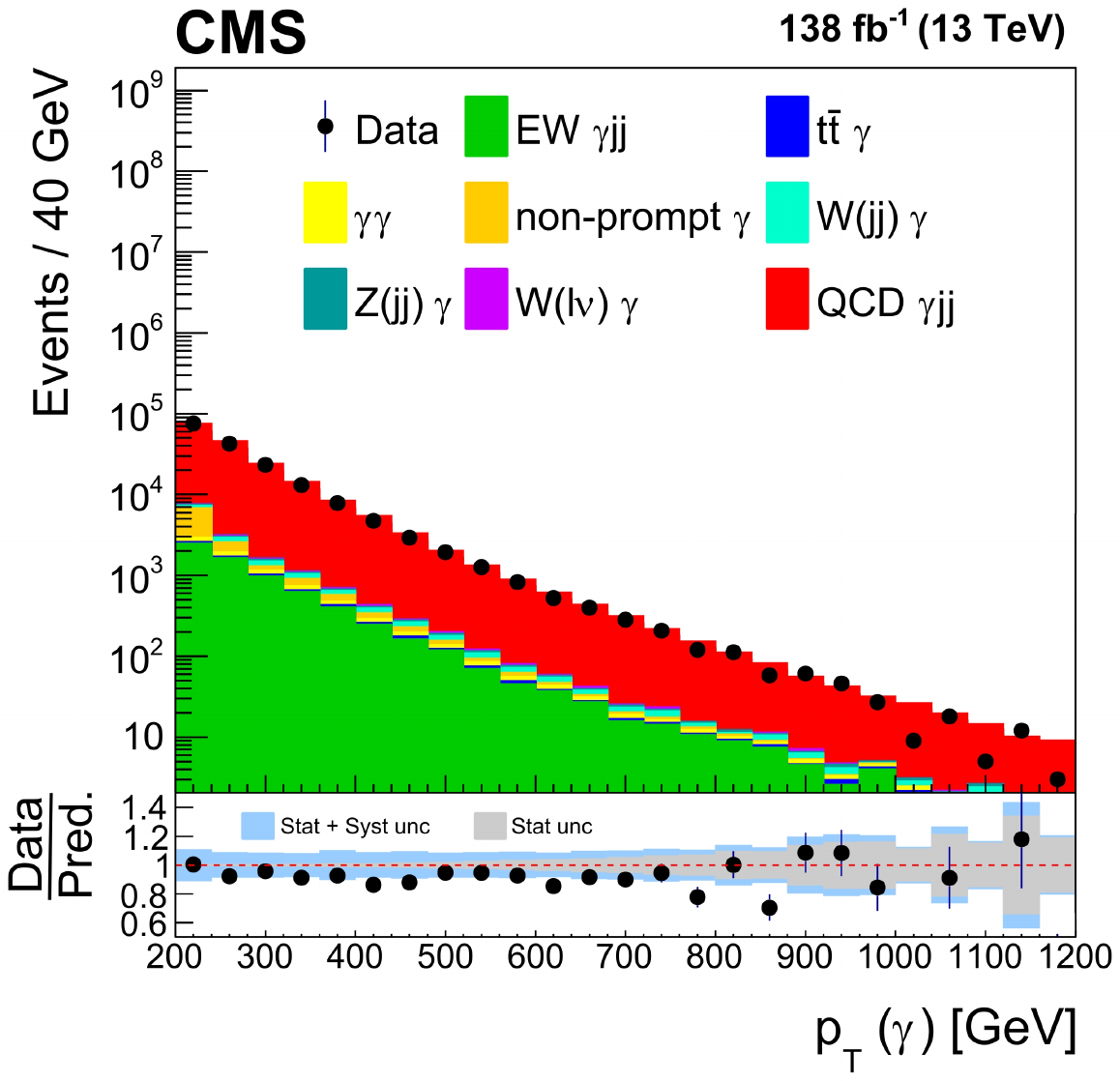}
      \includegraphics[width=0.49\textwidth]{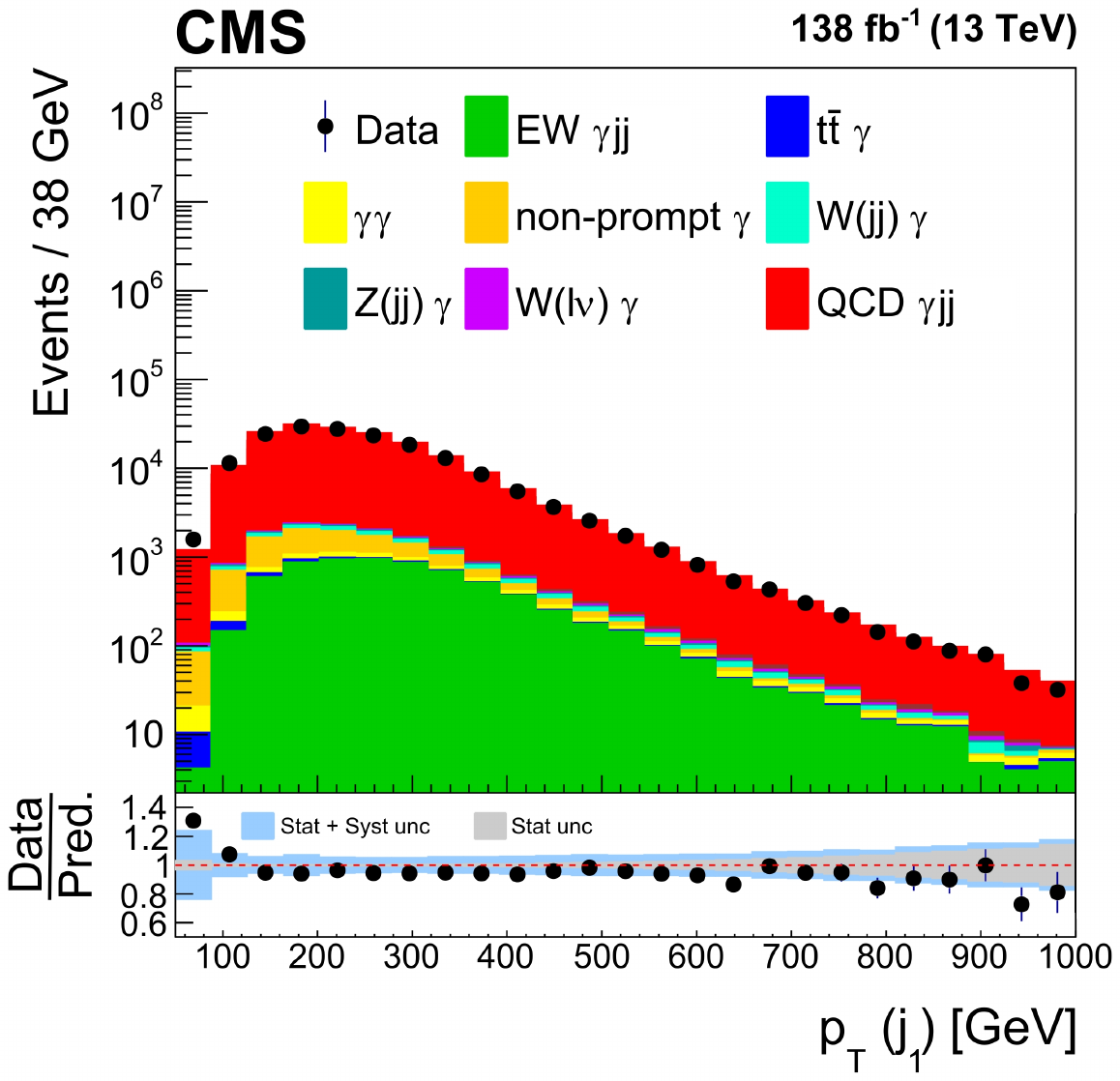}\\
      \includegraphics[width=0.49\textwidth]{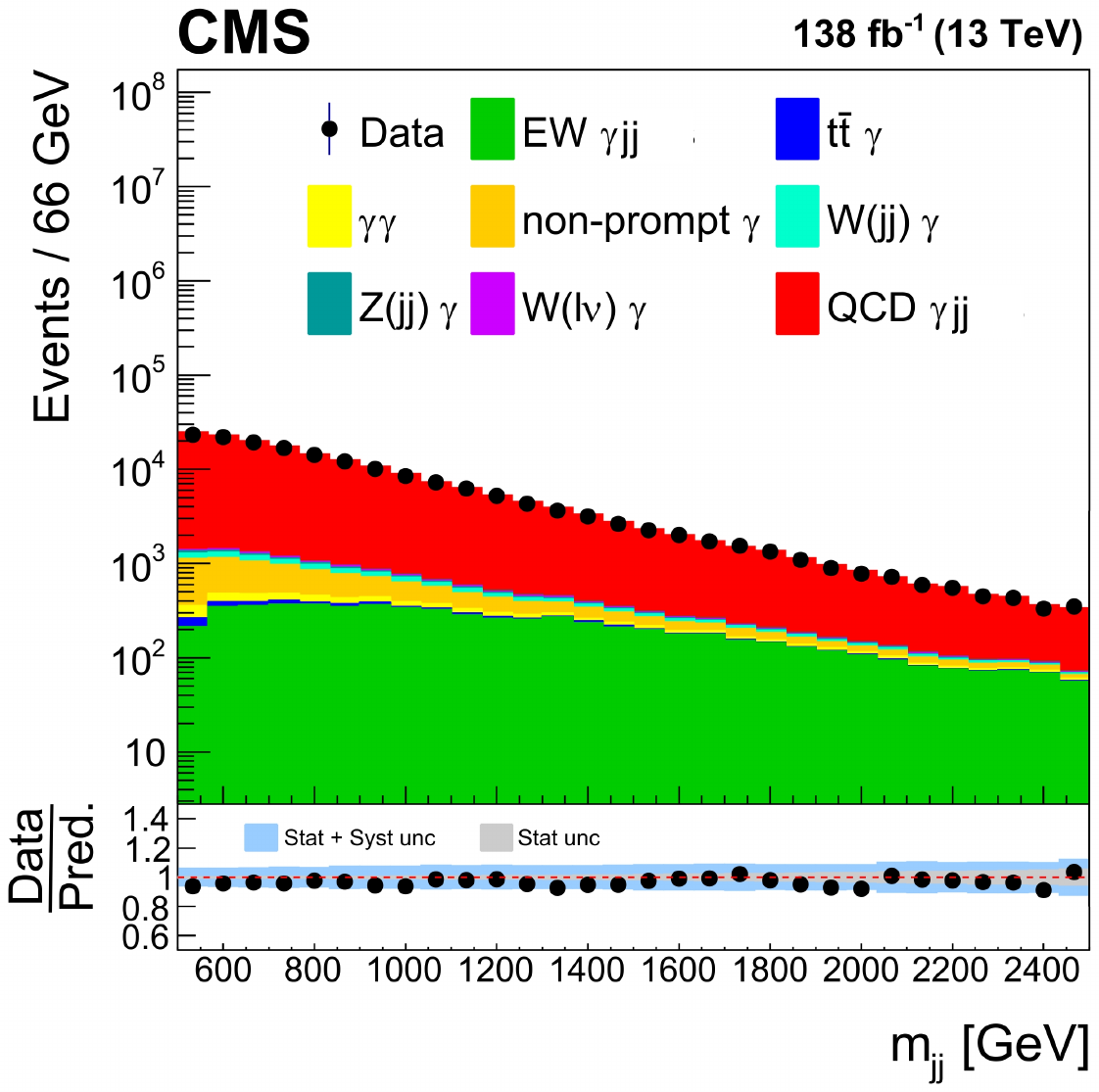}
      \includegraphics[width=0.49\textwidth]{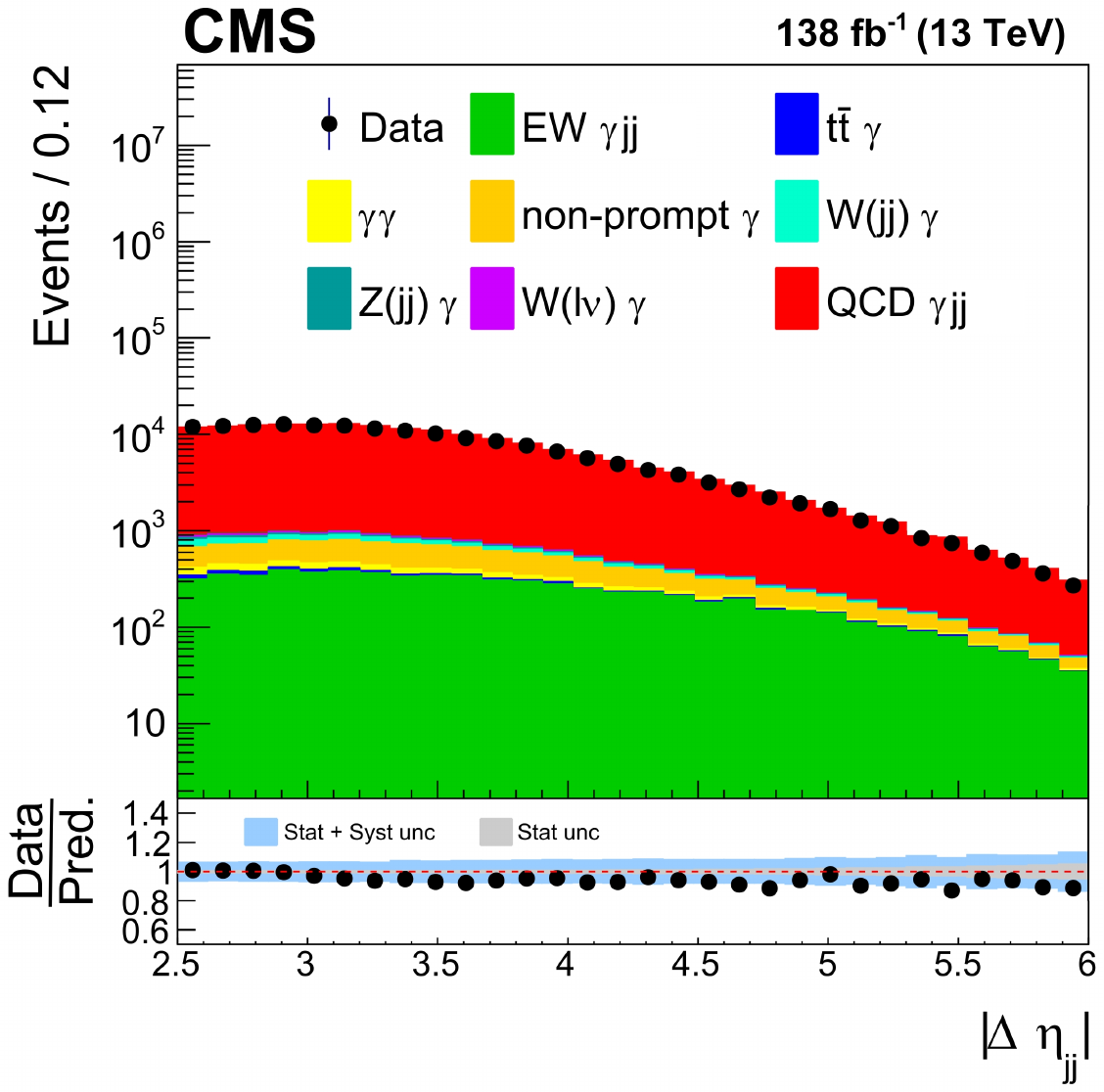}
      \caption{Distribution of (upper left) photon \pt, (upper right) leading jet \pt, (lower left) \mjj, and (lower right) $\abs{\detajj}$ in data and simulated processes, except the contribution of nonprompt photons that is estimated from data as discussed in Section~\ref{bkg_estimation}. Simulted samples are normalized to their theoretical cross sections. The black points with error bars represent the data and their statistical uncertainties. The last bin includes the overflow events. The lower panels shows the ratio of the data to the expectation with the inner (outer) band representing the statistical (total) uncertainty in the combined signal and background expectations.}
      \label{fig:prefit}
\end{figure}

\begin{figure}[ht!]
   \centering
      \includegraphics[width=0.49\textwidth]{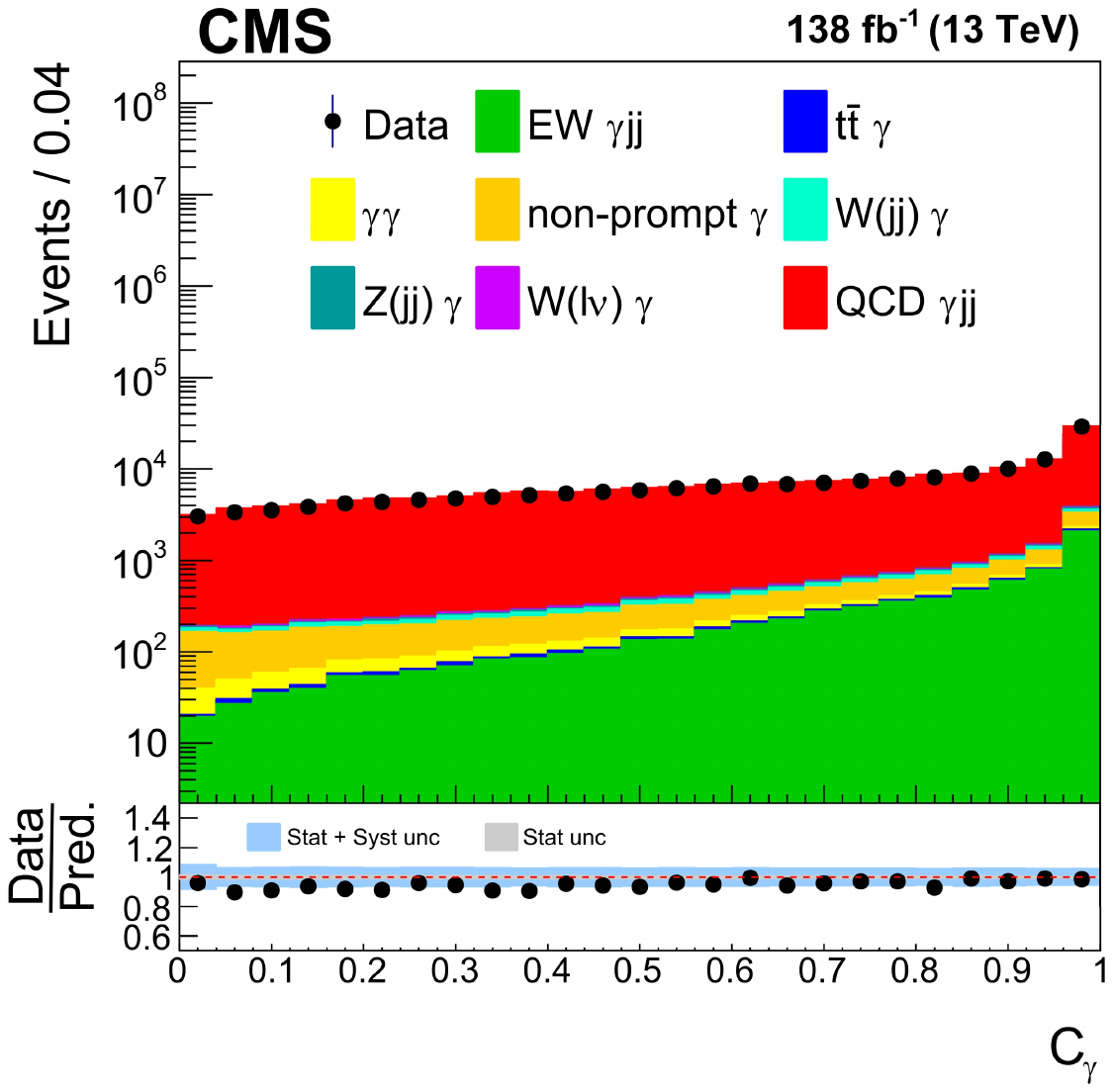}
      \includegraphics[width=0.49\textwidth]{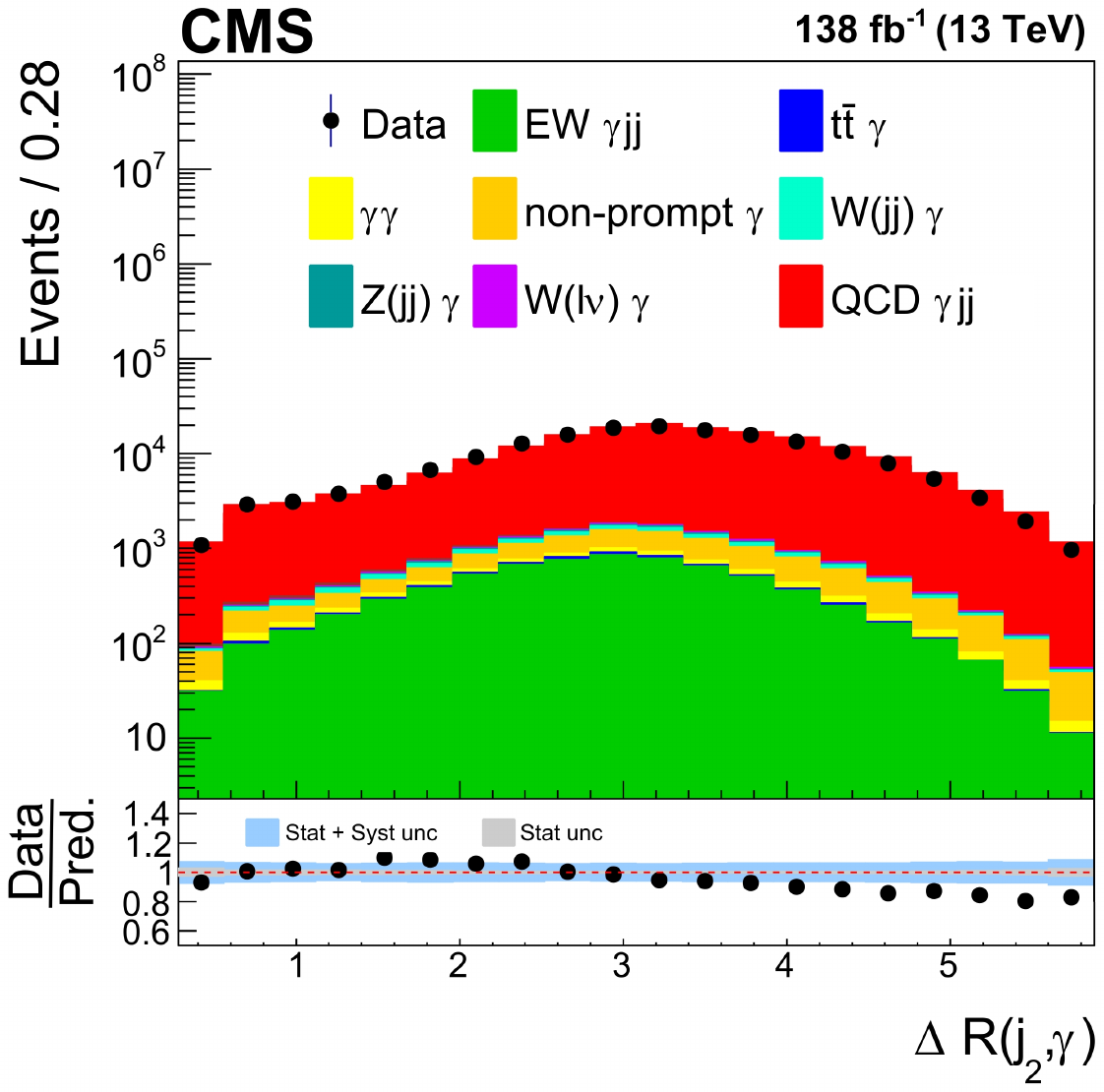}
      \includegraphics[width=0.49\textwidth]{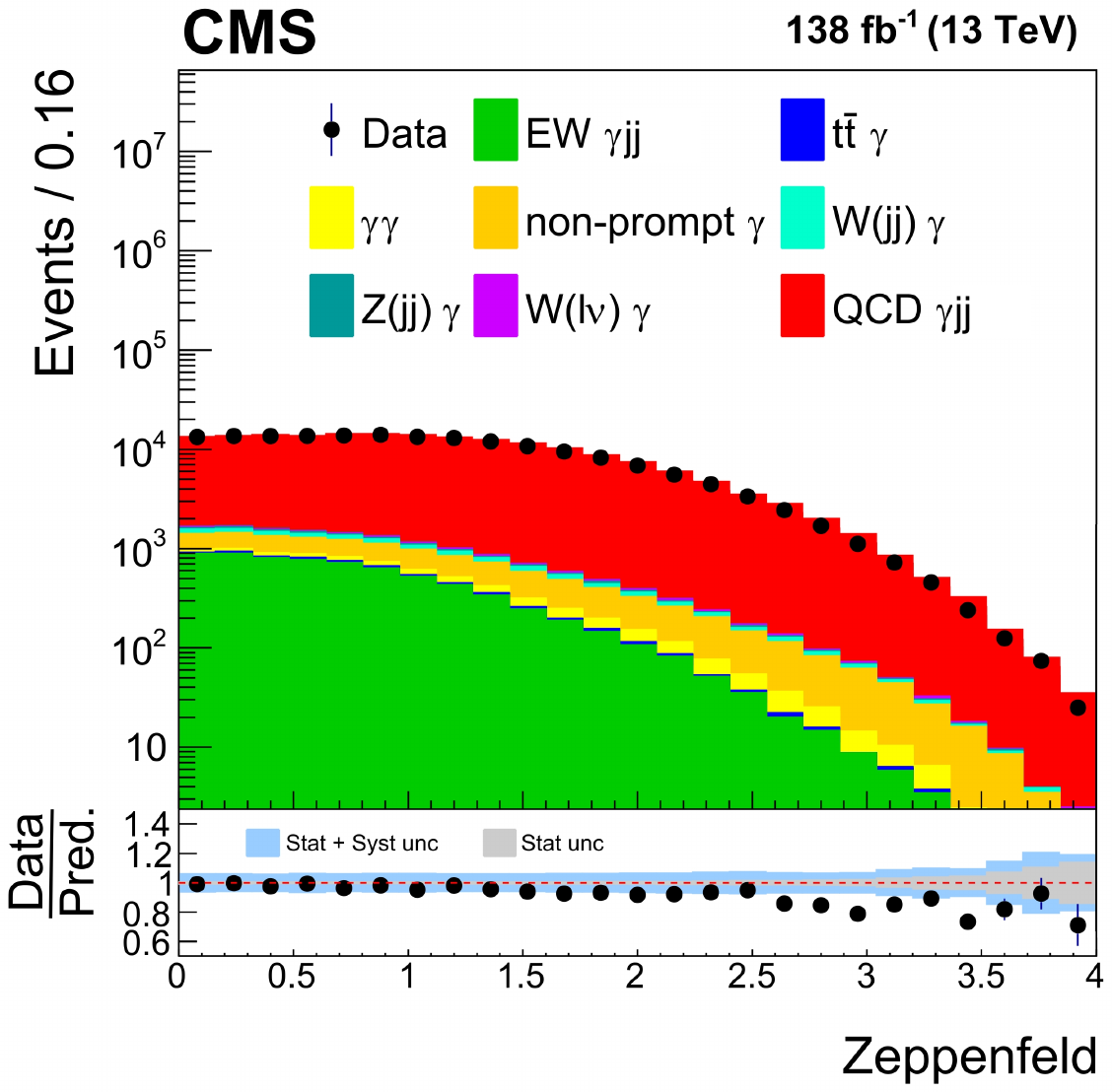}
      \caption{Distribution of (upper left) $C_{\PGg}$, (upper right) $\Delta R(j_2,\PGg)$, and (lower) the Zeppenfeld variable in data and simulated processes, except the contribution of nonprompt photons that is estimated from data as discussed in Section~\ref{bkg_estimation}. Simulted samples are normalized to their theoretical cross sections. The black points with error bars represent the data and their statistical uncertainties. The last bin includes the overflow events. The lower panels shows the ratio of the data to the expectation with the inner (outer) band representing the statistical (total) uncertainty in the combined signal and background expectations.}
      \label{fig:bdtinput}
\end{figure}

Figures~\ref{fig:prefit} and~\ref{fig:bdtinput} show the distributions of some of these variables where, in general, the data are well described by the expected distributions within uncertainties. Minor discrepancies are observed in the tails of some of the distributions that could be attributed to the mismodeling of QCD \ajj at NLO. A binned maximum likelihood fit of the BDT distribution is performed to the data with systematic uncertainties included as nuisance parameters using the CMS statistical analysis tool \textsc{Combine}~\cite{CMS:2024onh} to measure inclusive and differential EW \ajj cross sections. Events with a BDT score greater than 0.6 are used for the interpretation of the results in the context of an effective field theory. 

\section{Systematic uncertainties}
\label{uncertainties}
Systematic uncertainties with theoretical and experimental origins are considered in the measurements. The choice of renormalization and factorization scales, variations in parameters of the PDF model, and the scales of the parton shower are among the theory uncertainties. Experimental uncertainties are associated either with the analysis techniques, such as the estimation of the nonprompt photon contribution, or with corrections to simulation, including the kinematics of jets and photons, the identification and selection efficiencies, and the effect of pileup, as well as detector deficiencies during data taking that were not simulated. 

Uncertainties due to missing higher order corrections in QCD are estimated by varying the factorization and renormalization scales, \muf and \mur, respectively, by a factor of 2 or 0.5, both individually and simultaneously. The anticorrelated cases are excluded and the envelope of variations in the shape of the BDT distribution is considered as an uncertainty. 
     
Uncertainties arising from variations in the strong coupling constant $\alpS$ and PDFs are evaluated using the eigenvalues of the PDF set following the NNPDF prescription~\cite{Butterworth:2015oua}. Parton shower uncertainties arise from the scale variations in the ISR and FSR modeling. By varying either the ISR or FSR scale by a factor of 2 or 0.5, there are four combinations, and the largest variation is taken as the systematic uncertainty. These theoretical uncertainties are evaluated for the signal and QCD \ajj background. The effects are considered correlated across data-taking periods while uncorrelated between processes. 

Trigger efficiencies and their corrections are derived as described in Section~\ref{event_selection}. The trigger efficiencies in data are evaluated using jets instead of muon triggers, and the difference is taken as an uncertainty. Additional uncertainties arise from the statistical uncertainty in the control data and the limited size of the MC sample. Besides, imposing $\mjj>500\,\GeV$, corrections are recomputed and the difference is considered as uncertainty. Variations from all sources are added in quadrature, resulting in the  trigger uncertainty as a function of \ptg. The uncertainty is considered uncorrelated between different data-taking periods.

Uncertainties associated with the determination of the jet energy, \ie, the jet energy scale (JES) and resolution (JER), are evaluated in simulation and split into several uncorrelated sources~\cite{CMS:2016lmd}. Uncertainties are obtained by varying each source by $\pm 1$ standard deviation. Changes in jet momenta are propagated to BDT inputs constructed using jet kinematics. Variations in the BDT distribution are included in the fit and considered correlated between different processes. A correlation pattern is defined~\cite{CMS:2016lmd} across the data-taking periods, depending of the source.

Uncertainties in the photon identification efficiency and pixel seed veto are considered~\cite{CMS:2020uim}, affecting the BDT shape and the normalization of contributing processes. Variations are within 1--10\% depending on \pt and include both statistical and systematic components. The systematic contribution dominates at low \pt, whereas the statistical part becomes significant at higher \pt of the photon.

A problem with the ECAL triggers during 2016--2017 resulted in a small fraction of events selected from the previous proton bunch crossing~\cite{CMS:2024ppo}. Simulated events are corrected to account for this timing shift and corresponding uncertainties are assigned.

The modeling of nonprompt photons is affected by three uncertainty sources described in Section~\ref{bkg_estimation}, namely, the choice of the isolation sideband, the bias in the fitting procedure, and the shower shape of prompt photons. Summing in quadrature, the uncertainty in the estimation of nonprompt photons ranges from 25 to 90\%, depending on photon \pt. This uncertainty is considered uncorrelated across the years of data taking.

Backgrounds with prompt photons are dominated by the QCD \ajj production, whose normalization is unconstrained and implemented as a free normalization parameter in the fit. The minor contribution of other backgrounds, including also the interference between the EW and QCD \ajj, is assigned 20\% uncertainty, covering possible mismodeling in the phase space of the analysis. This uncertainty is considered correlated between data-taking periods.

Uncertainties associated with the integrated luminosity for different data-taking periods are determined to be within 1.2--2.5\%~\cite{CMS-LUM-17-003, CMS:2018elu,CMS:2019jhq}. They are considered partially correlated across years of data taking, resulting in an overall uncertainty for 2016--2018 of 1.6\%. The uncertainty arising from pileup is evaluated varying the total inelastic cross section by 4.6\% and is considered correlated between years~\cite{minibais}. The finite number of simulated events introduces an uncertainty that is accounted for using the Beeston-Barlow ``lite'' method~\cite{Barlow:1993dm}. 

\section{Inclusive and differential measurements} 
\label{significance}
\begin{figure*}[!htbp]
   \centering
      \includegraphics[width=0.45\textwidth]{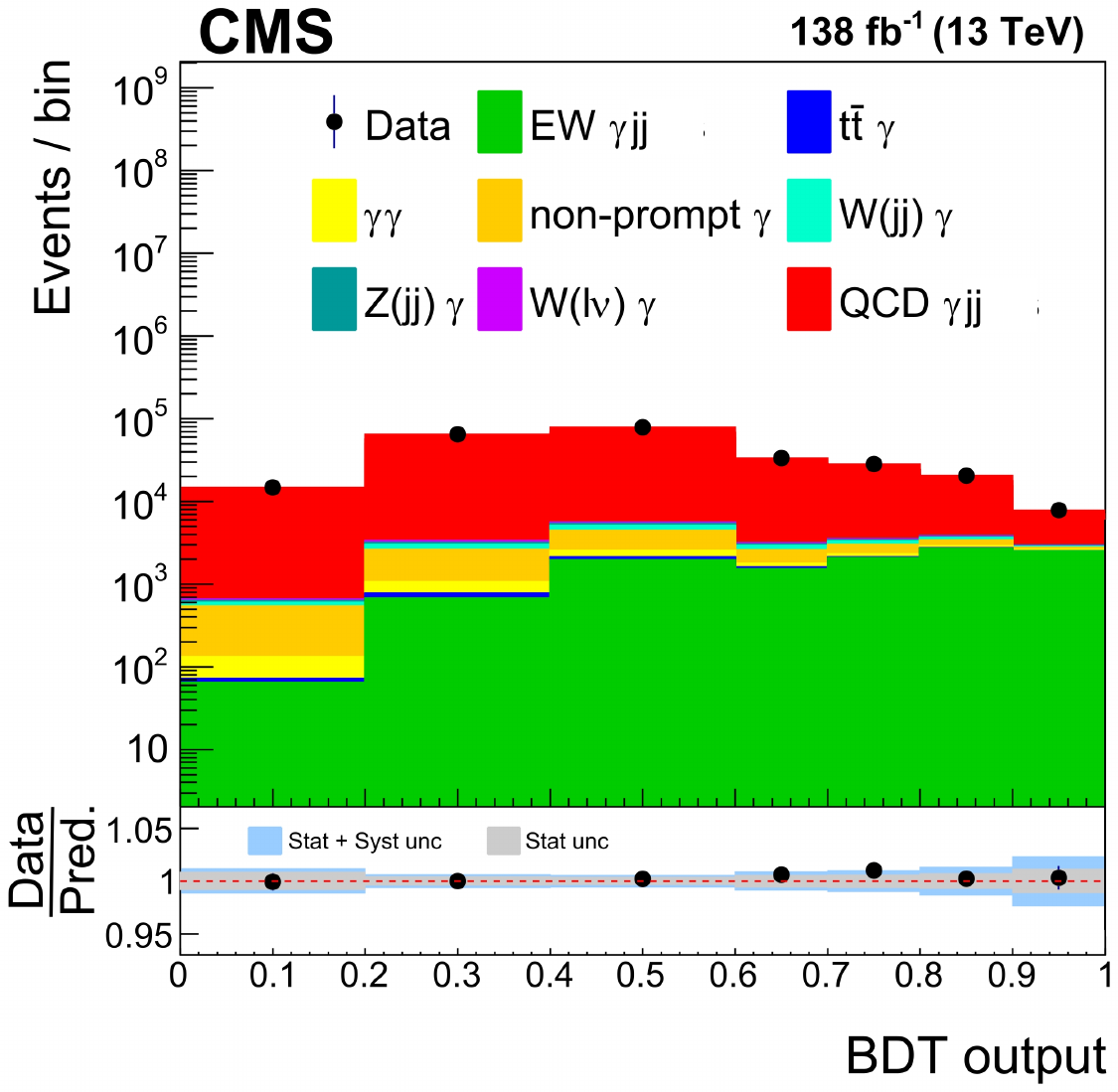}
      \caption{The postfit BDT output distribution. The data are compared to the sum of the signal and the background contributions. The black points with error bars represent the data and their statistical uncertainties. The lower panel shows the ratio of the data to prediction where the inner (outer) band represents the statistical (total) uncertainty in the combined signal and background contributions after the fit. }
      \label{fig:postfit_plots}
\end{figure*}

The normalization of the QCD \ajj background, unconstrained in the fit to extract the inclusive $\ewsig$, is determined to be $1.00 \pm 0.06$ relative to the expected value. The EW \ajj signal is observed over the expected backgrounds with a significance of more than five standard deviations, compatible with the expected significance from simulation. The measured fiducial cross section is \xsec{}, in agreement with the SM prediction $177^{+13}_{-12}~\mathrm{fb}$, where the single dominant systematic uncertainty is associated with the jet energy scale and resolution. The BDT distribution after the fit to data is shown in Fig.~\ref{fig:postfit_plots}. A summary of systematic uncertainties after the fit is provided in Table \ref{tab:unc}.

\begin{table}[ht]
    \centering
    \renewcommand{\arraystretch}{1.2}
    \caption{Summary of uncertainties affecting the measurement as extracted from the fit to data. The total uncertainty is obtained by adding individual contributions in quadrature.}
    \label{tab:unc}
    \begin{tabular}{lc}
	    Source & Uncertainty (fb) \\
        \hline
        Data statistical uncertainty & $\pm$7 \\[\cmsTabSkip]
        Limited size of MC samples & $\pm$4 \\
        Normalization of minor backgrounds  & $\pm$6 \\
        Theory      & $+14,-12$ \\
        Jet energy scale and resolution & $\pm$22 \\
        Other experimental uncertainties & $+22,-20$ \\[\cmsTabSkip]
        Total & $+36,-32$ \\
    \end{tabular}
\end{table}

\begin{figure*}[htbp]
   \centering
      \includegraphics[width=0.5\textwidth]{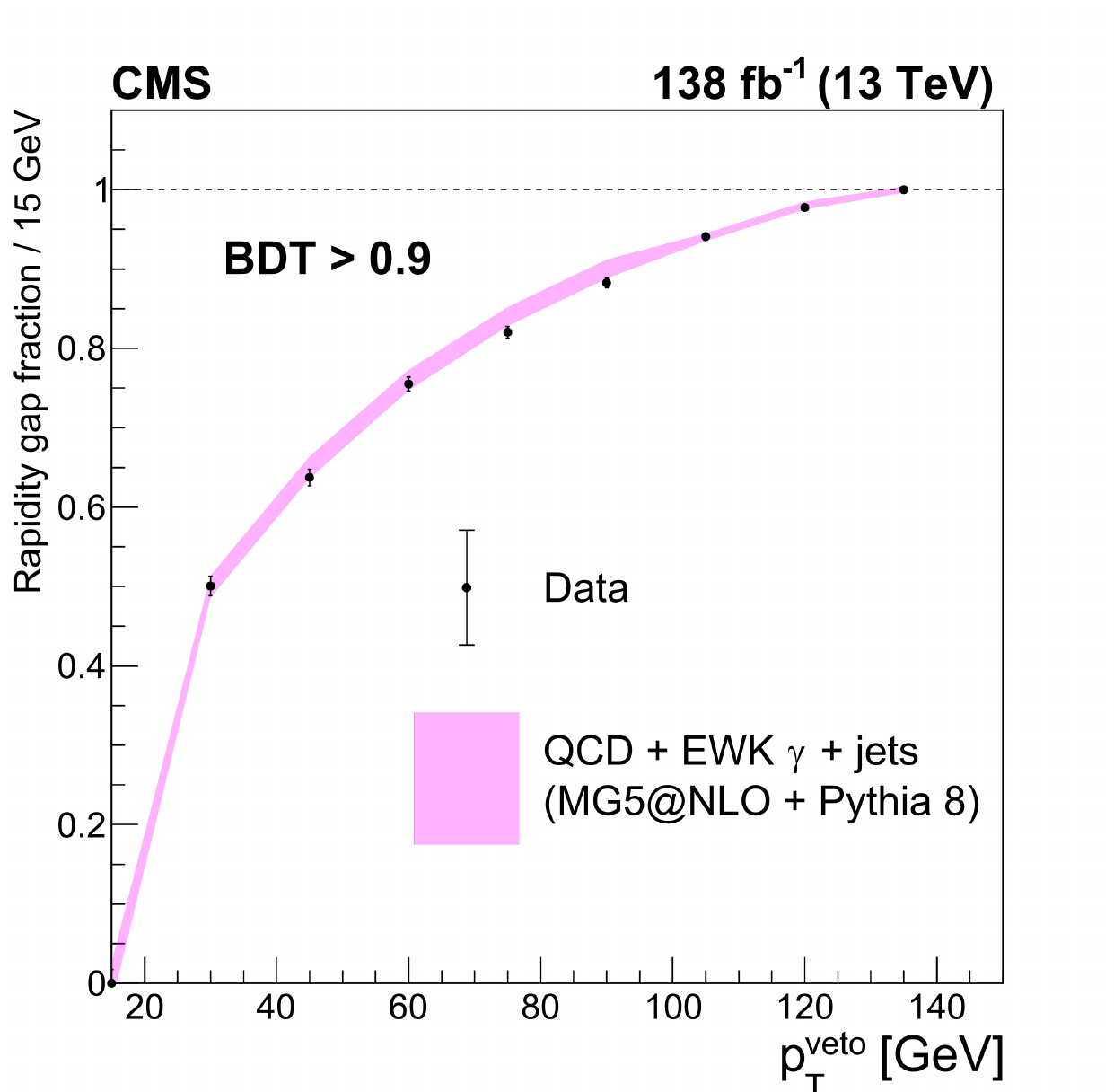}
      \caption{The rapidity gap fraction as a function of $\pveto$ in data and simulated samples for EW \ajj and QCD \ajj. The black points with error bars represent the data and their statistical uncertainties. The theory prediction, calculated using MG5+\PYTHIA, together with the MC statistical uncertainties are shown by the colored band. }
      \label{fig:gap}
\end{figure*}

The hadronic activity in the pseudorapidity gap between the two tagging jet is studied using gap jets. In a pure VBF process, jets in the pseudorapidity gap are expected to come from parton showers, hence carrying low energies. 
Considering an upper bound on the gap jet transverse momentum, $\pveto$, the rapidity gap fraction is defined as the fraction of events in which the gap jet satisfies $ \pt < \pveto$. 
In order to enrich the data sample in VBF EW \ajj, events are required to have a BDT score greater than 0.9. The contribution of the EW \ajj signal is about 24\% in the selected sample. The rapidity gap fraction as a function of $\pveto$ is shown in Fig.~\ref{fig:gap} for data and the combination of EW \ajj and QCD \ajj in simulation. Expected event yields are scaled using the fit results of the inclusive EW \ajj cross section measurement. Other backgrounds are negligible and are, therefore, not considered at this stage. The data are well described by simulation, indicating a reasonable modeling of the jet activity in \PYTHIA~8.240.

\begin{figure*}[htbp]
   \centering
      \includegraphics[width=0.95\textwidth]{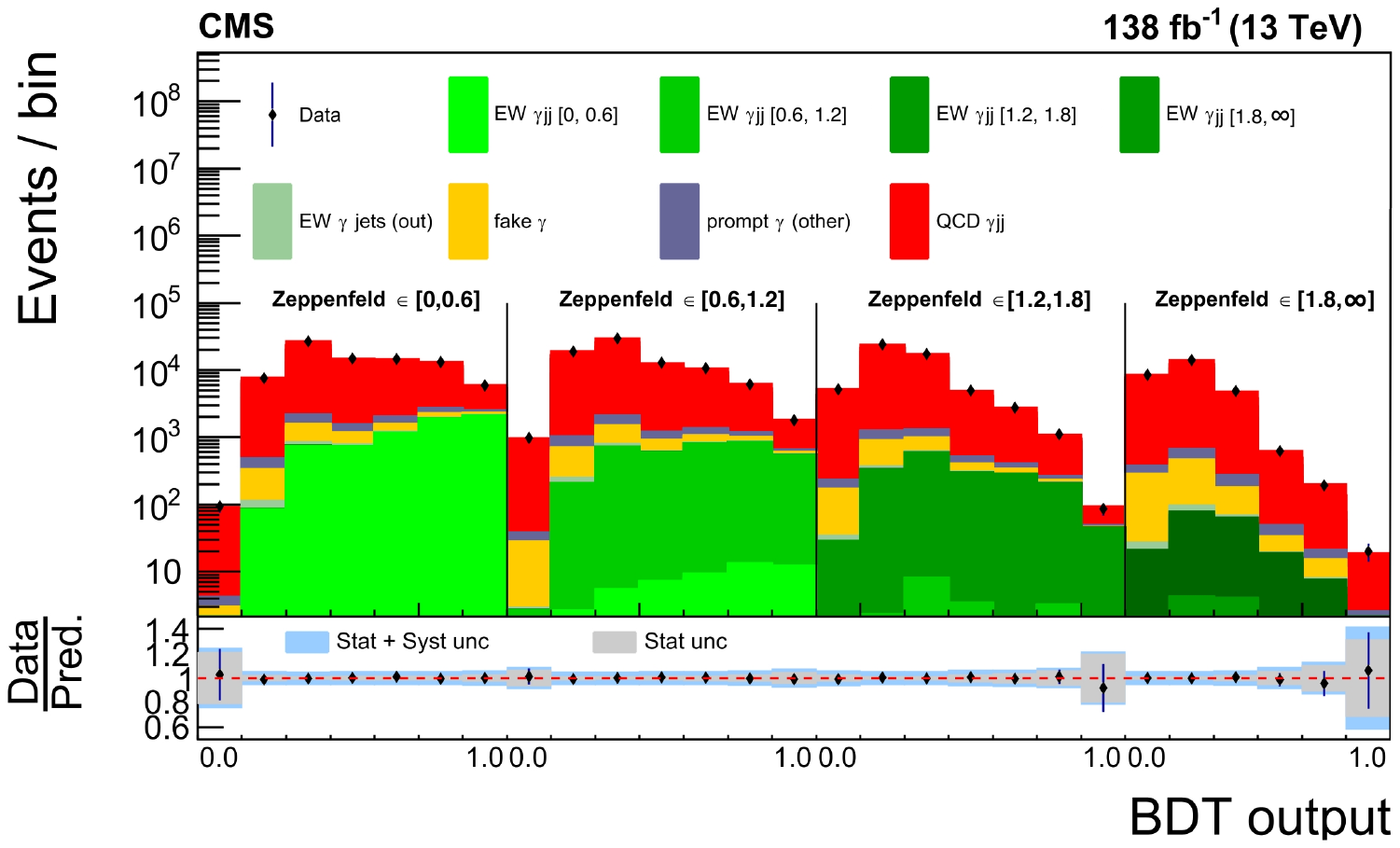}
	\caption{The unrolled BDT distribution in bins of the Zeppenfeld observable after the fit to the data. Signal events from different Zeppenfeld ranges at the generator level are represented by different colors, whereas different Zeppenfeld ranges at the detector level are displayed as an overlaid distribution. The different shades of green correspond to increasing ranges of the Zeppenfeld observable at the generator level ([0,0.6],[0.6,1.2],[1.2,1.8],[1.8,$\infty$]). The label ``out'' refers to signal events outside the defined phase space. The black points with error bars represent the data and their statistical uncertainties. The lower panel shows the ratio of the data to the prediction. The inner and outer bands represent, respectively, the statistical and total uncertainties on all simulated samples after the fit. }
      \label{fig:2dmjj}
\end{figure*}

Differential $\ewsig$ measurements are unfolded back to the particle level as functions of observables mainly describing the VBF signature of the signal. A likelihood-based unfolding procedure is followed where the signal yield in every bin of an observable at the detector level is parameterized as a function of EW \ajj cross sections in all bins of the same observable at the particle level. A simultaneous fit in all bins of the observable is performed to the data using the BDT distribution with all systematic uncertainties included as nuisance parameters. The use of the BDT distribution preserves the sensitivity to the signal in the presence of the large QCD \ajj background. Figure~\ref{fig:2dmjj} shows the unrolled BDT distribution in bins of the Zeppenfeld observable after the fit to the data. Signal events from different Zeppenfeld ranges at the generator level are represented by different colors. 
 
\begin{figure*}[htbp]
   \centering
      \includegraphics[width=0.44\textwidth]{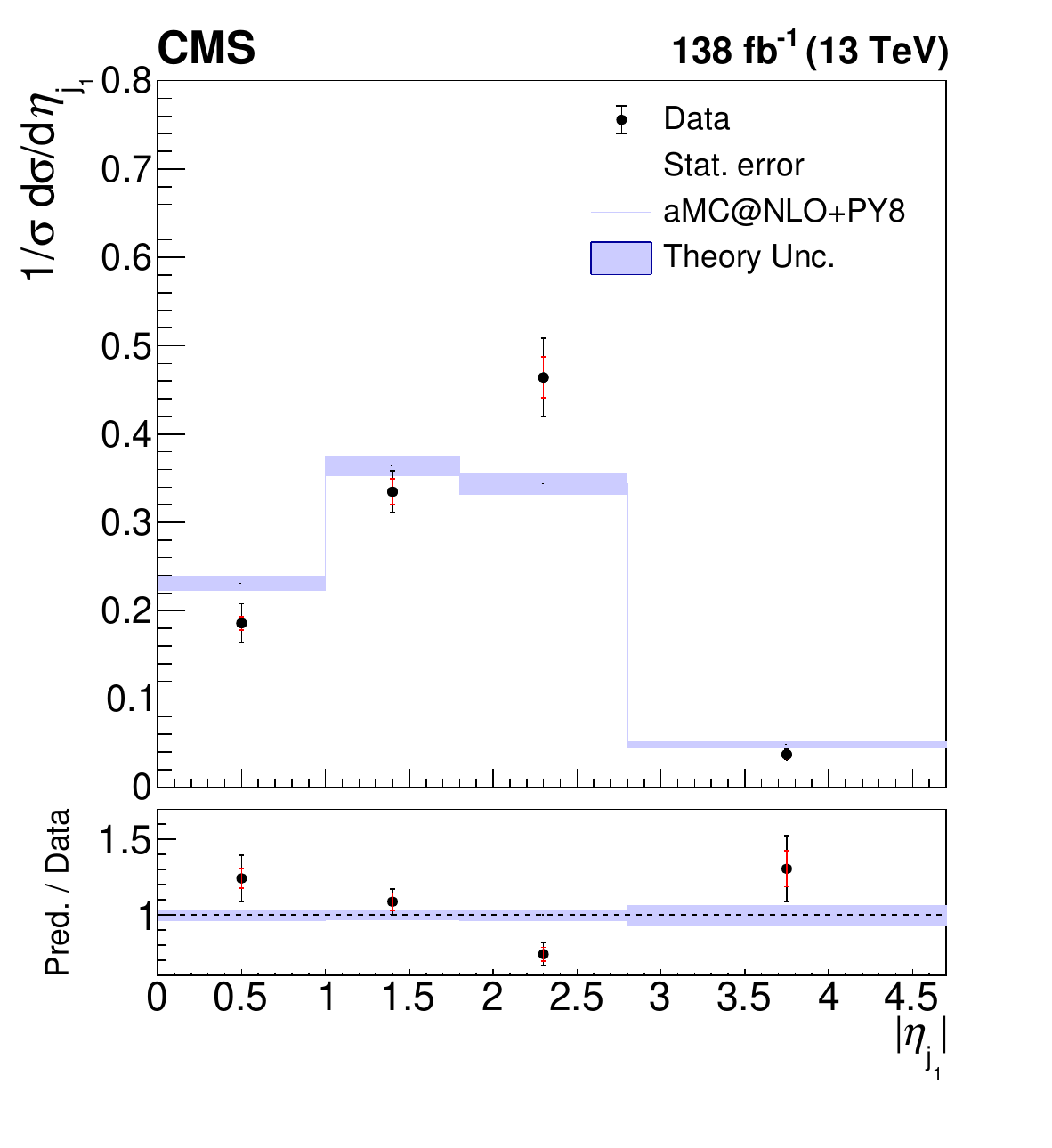}
      \includegraphics[width=0.44\textwidth]{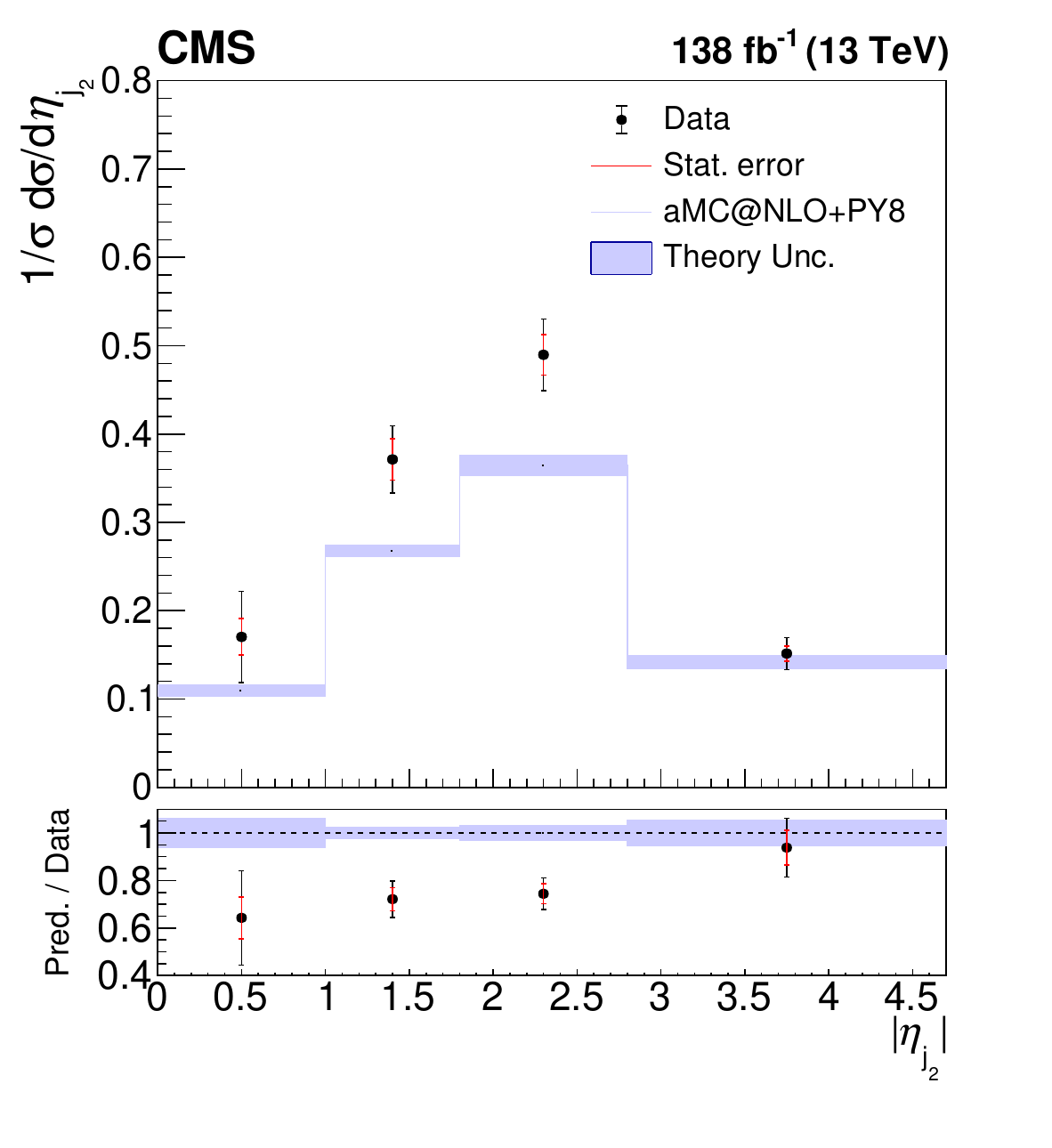}\\
      \includegraphics[width=0.44\textwidth]{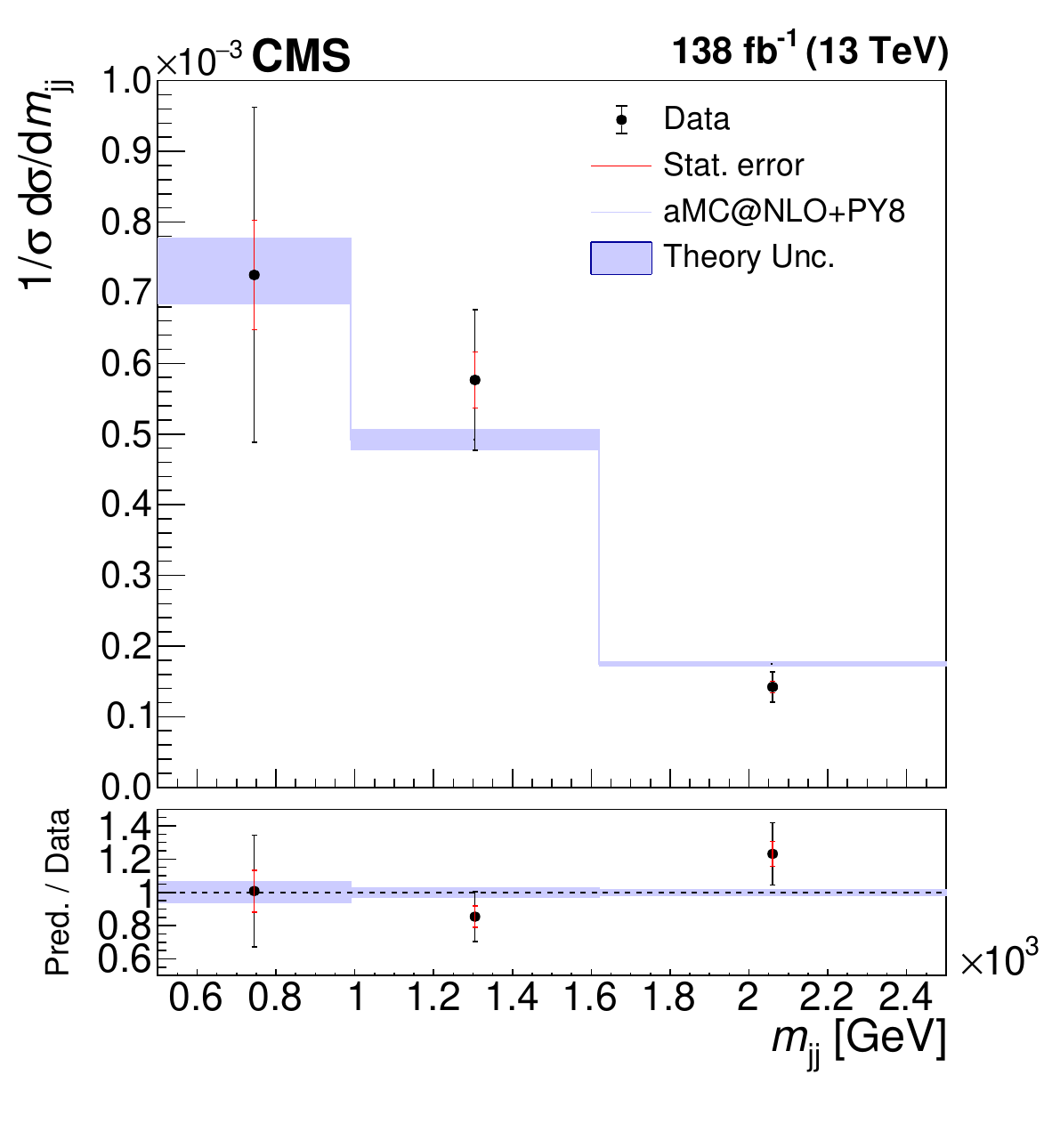}
      \includegraphics[width=0.44\textwidth]{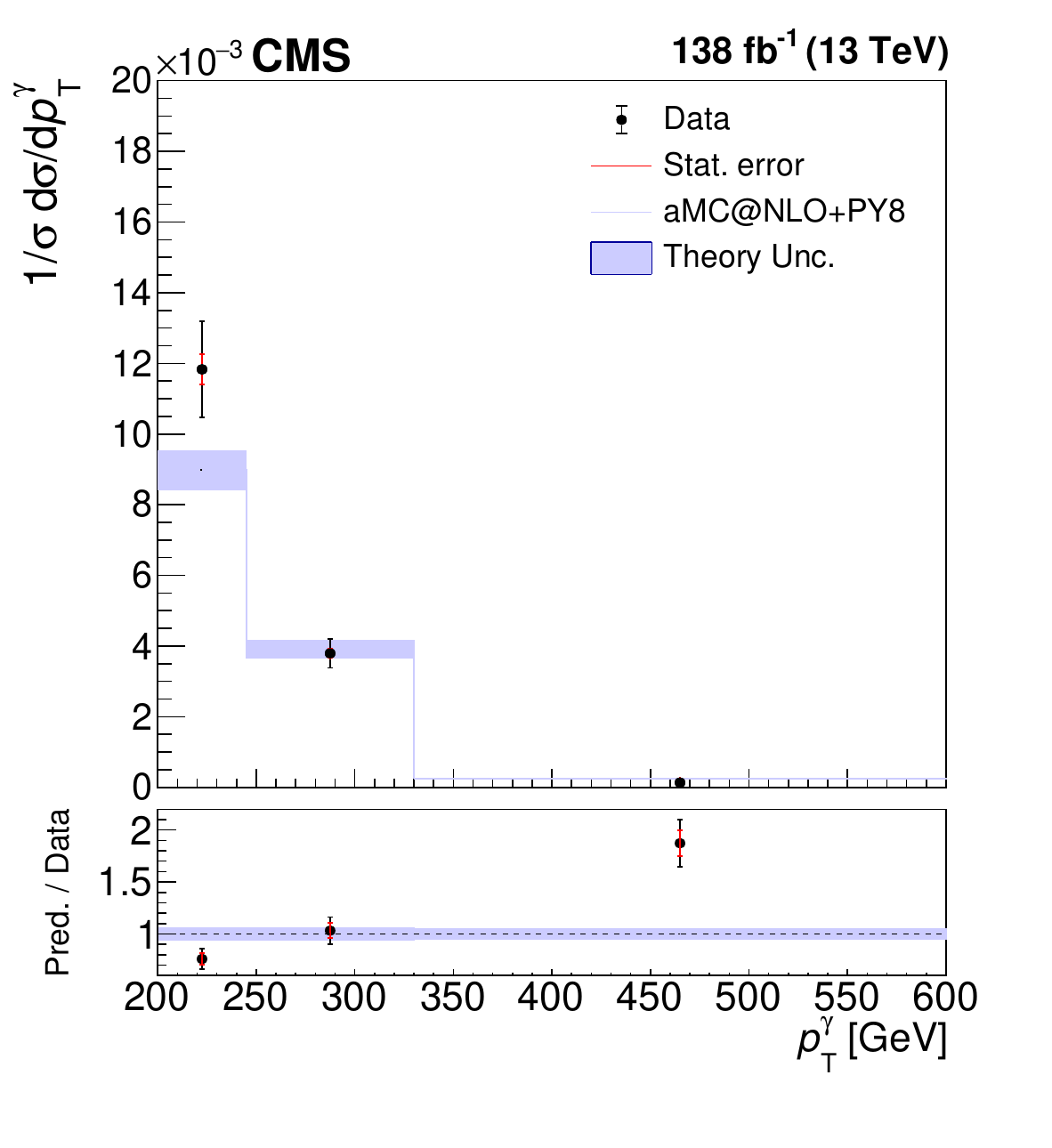}\\
      \includegraphics[width=0.44\textwidth]{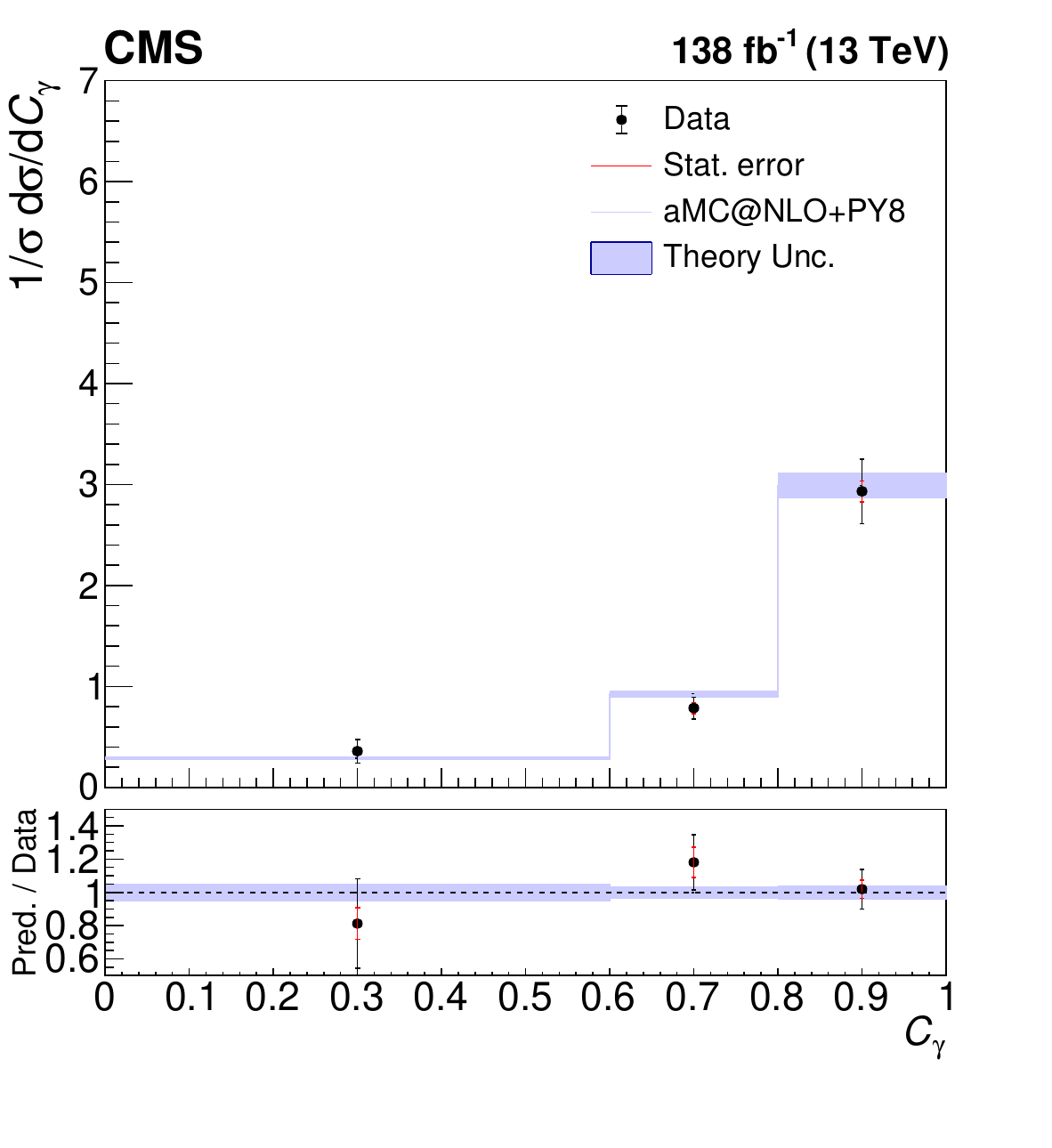}
      \includegraphics[width=0.44\textwidth]{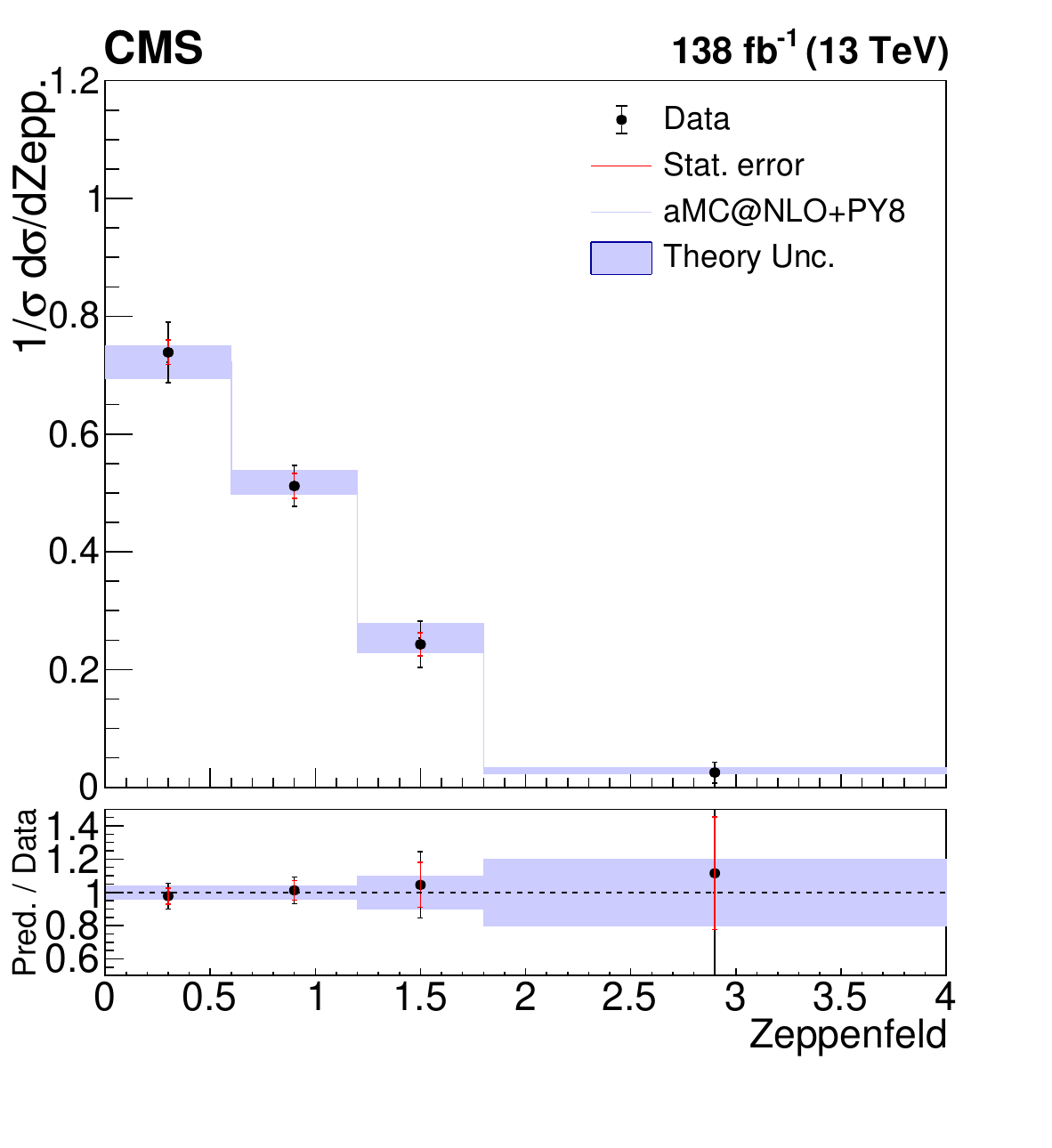}
      \caption{Normalized differential cross sections, compared with the SM predictions, as functions of (upper left) \etajl, (upper right) \etajs, (middle left) \mjj, (middle right) $\pt^\PGg$, (lower left) \cg, and (lower right) the Zeppenfeld variable. The red bars on the data points represent the statistical errors, whereas the black bars show the total uncertainties.}
      \label{fig:unfold_results}
\end{figure*}

The normalized differential cross sections are directly extracted from the fit by introducing, in bin $i$ of the differential distribution, the ratio between $\sigma^i_{\text{EW}\ajj}$ and the total fiducial $\ewsig$ as a parameter of interest, while otherwise the fits are structured identically to the un-normalized differential fit. Allowing $\ewsig$ to vary in the fit, the normalized cross section of one of the bins in the differential distribution is parameterized as a function of the remining bins. As a result, the normalized cross sections retain the same number of degrees of freedom as a standard differential measurement. This approach significantly reduces the systematic uncertainties and accounts for correlations by construction. The normalized differential cross sections are shown, as functions of $\ptg$, \etajl, \etajs, \mjj, \cg, and the Zeppenfeld variable, in Fig.~\ref{fig:unfold_results}, compared with predictions. Uncertainties in the predictions are estimated by summing in quadrature the systematic uncertainties coming from the renormalization and factorization scales, PDFs, and parton shower simulation. 

The measurements are in agreement with predictions across all observables, except for a few bins, particularly in the $\etajs$ distribution. It can be due to the dominant QCD $\PGg$jj background, where the second jet is not included in the matrix-element calculation but instead modeled through hadronization. This treatment can lead to a mismodeling of the distribution in the signal region. 

\section{The EFT interpretation}
\label{eft}

The sensitivity of the analysis to new phenomena in the $\PW\PW\PGg$ interaction is examined in SMEFT~\cite{Ellis:2021kzk}. The Wilson coefficients (WCs) of the dimension-6 operators, $\cw$ and $\chwb$, are considered for the study as they are the two operators that genuinely modify the $\PW\PW\PGg$ vertex~\cite{Brivio:2017vri}. The implementation of the EFT effects relies on the reweighting feature of MG5. To extract the weights, the \textsc{SMEFTsim}~3.0~\cite{Brivio:2020onw} model is employed with nonzero values of $\cw$ and $\chwb$. The real emissions at NLO in QCD are accounted for by including the production of the photon in addition to up to three jets. The EFT weights are then applied to the signal sample that is generated at NLO in QCD using \MGvATNLO. Several observables are studied to distinguish the SMEFT from the SM EW \ajj where for every observable, the proper MC modeling of background is ensured in comparison with the data. No single observable was found to provide an enough discriminating power for both $\cw$ and $\chwb$.

Considering the EW \ajj cross section within SMEFT, a deep neural network (DNN) is constructed to discriminate the pure SM part from the linear and quadratic terms. The training is performed using \textsc{Keras}~\cite{chollet2015keras} within the $\textsc{TensorFlow}$ platform~\cite{tensorflow2015-whitepaper}, providing enough robustness against negative event weights. The non-SM contributions to the cross section include the interference between the SM and EFT terms, the interference between the EFT operators corresponding to $\cw$ and $\chwb$, and the pure quadratic contributions of either of the EFT operators. The training is performed, simultaneously, against three different points in the ($\cw$, $\chwb$) parameter space, namely $(1,-1)$, $(1,0.4)$, and $(-1,-1)$, where two extreme and one intermediate scenario are considered. The training points are chosen such that, considering negative event weights, the effective size of each sample is statistically sufficient. Input features to the DNN include photon \pt, pseudorapidities of the leading and subleading jets, $\mjj$, and the $\Delta R$ between the photon and the subleading jet. The use of a DNN is further motivated by its improved handling of negative weights, which can become sizable in EFT analyses. The DNN output provides sensitivity to both operators, suitable to constrain them simultaneously, while surpassing the performance of individual observables.

The DNN distribution serves as an input to the statistical EFT analysis. The bin contents are parameterized according to different contributions to the SMEFT expected yield. Each event weight is a quadratic function of the WC, and bin yields are obtained by summing these quadratic weights over all events in the bin. This allows for an efficient parametrization of expected event yields as functions of EFT parameters. To increase the sensitivity of the analysis, events are required to have a BDT score of above 0.6, providing a region with a higher purity in EW \ajj.

\begin{figure*}[htbp]
   \centering
      \includegraphics[width=0.49\textwidth]{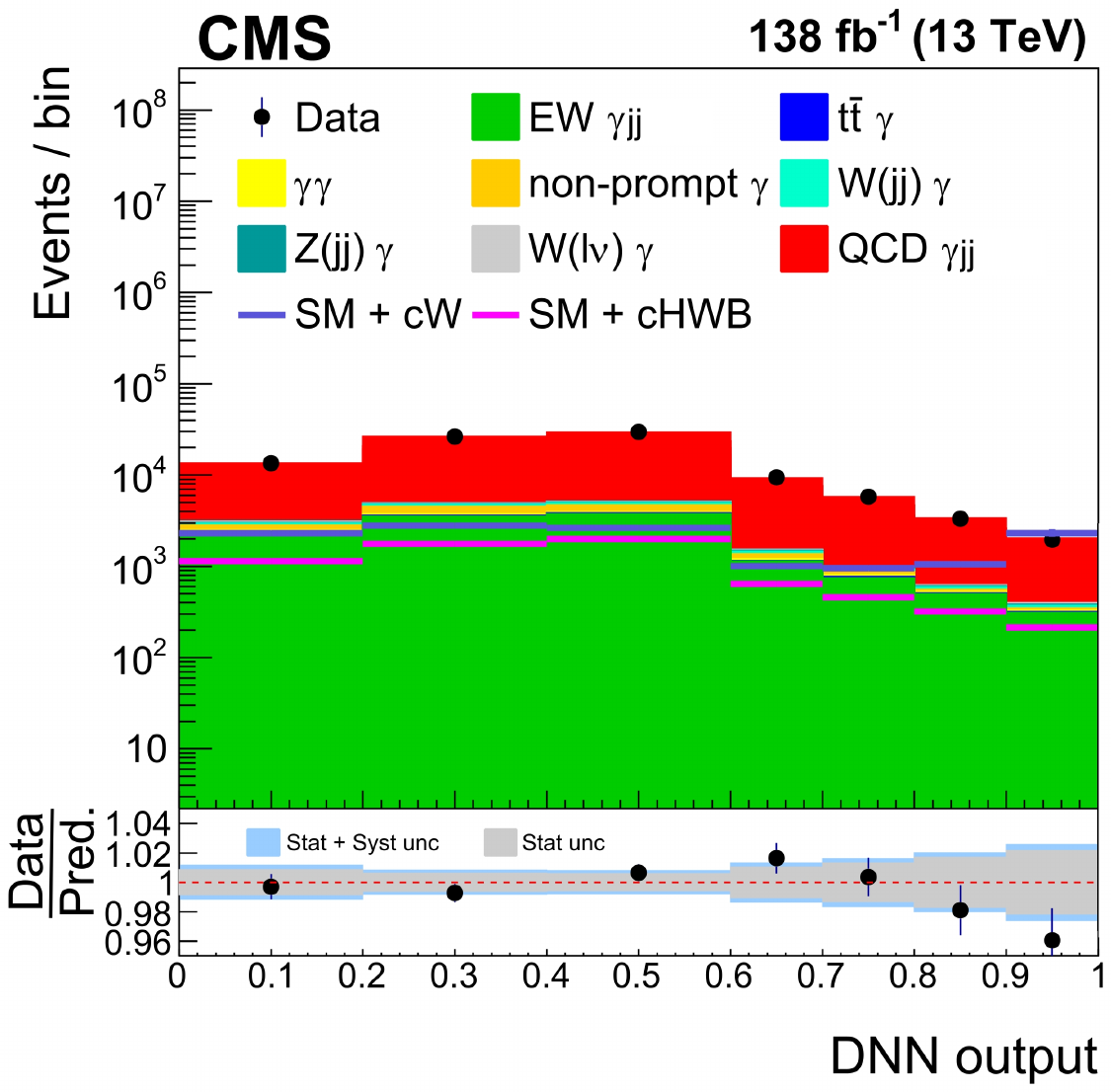}
      \caption{The distribution of the DNN output trained for \cw and \chwb coefficients in data and simulation. The simulation is corrected using the results of the inclusive \ewsig measurement. The black points with error bars represent the data and their statistical uncertainties. The purple and indigo lines show the distributions for the EW \ajj process when \chwb and \cw, respectively, are set to one. The lower panel shows the ratio of the data to the prediction. The inner and outer bands represent, respectively, the statistical and total uncertainties on all simulated samples as evaluated in the inclusive \ewsig measurement.}
      \label{fig:dnn}
\end{figure*}

\begin{figure*}[htbp]
   \centering
      \includegraphics[width=0.49\textwidth]{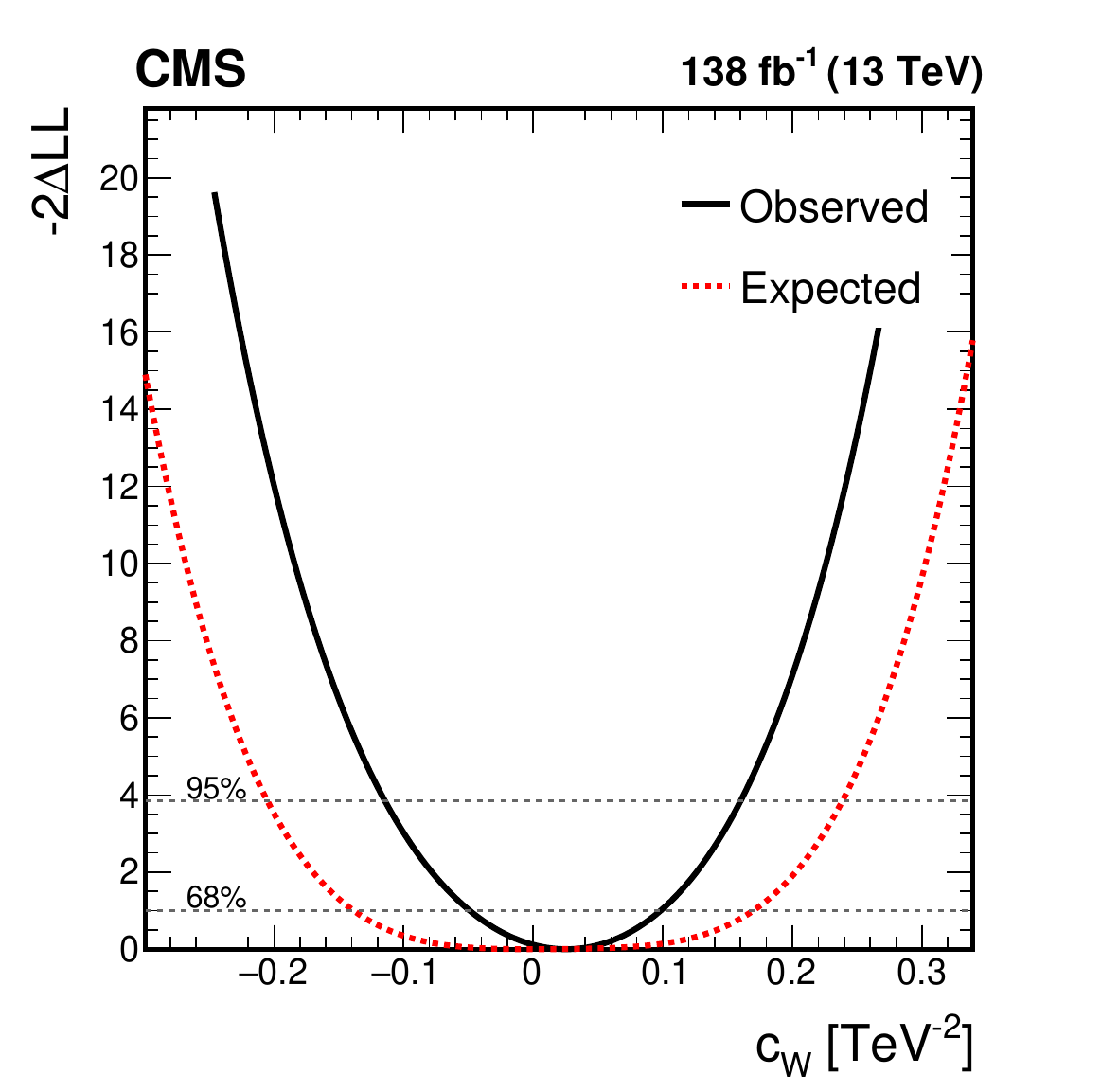}
      \includegraphics[width=0.49\textwidth]{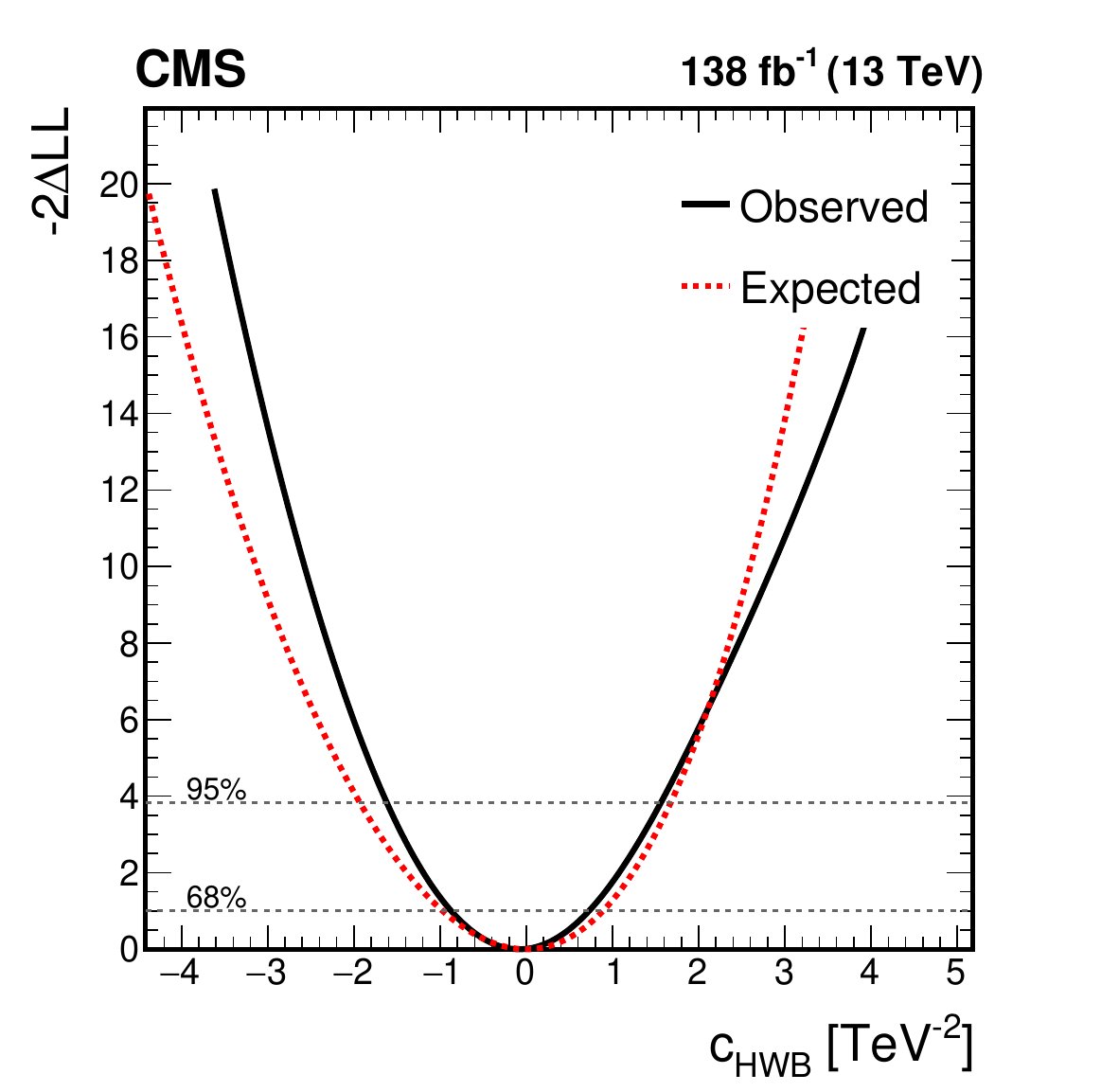}\\
      \includegraphics[width=0.49\textwidth]{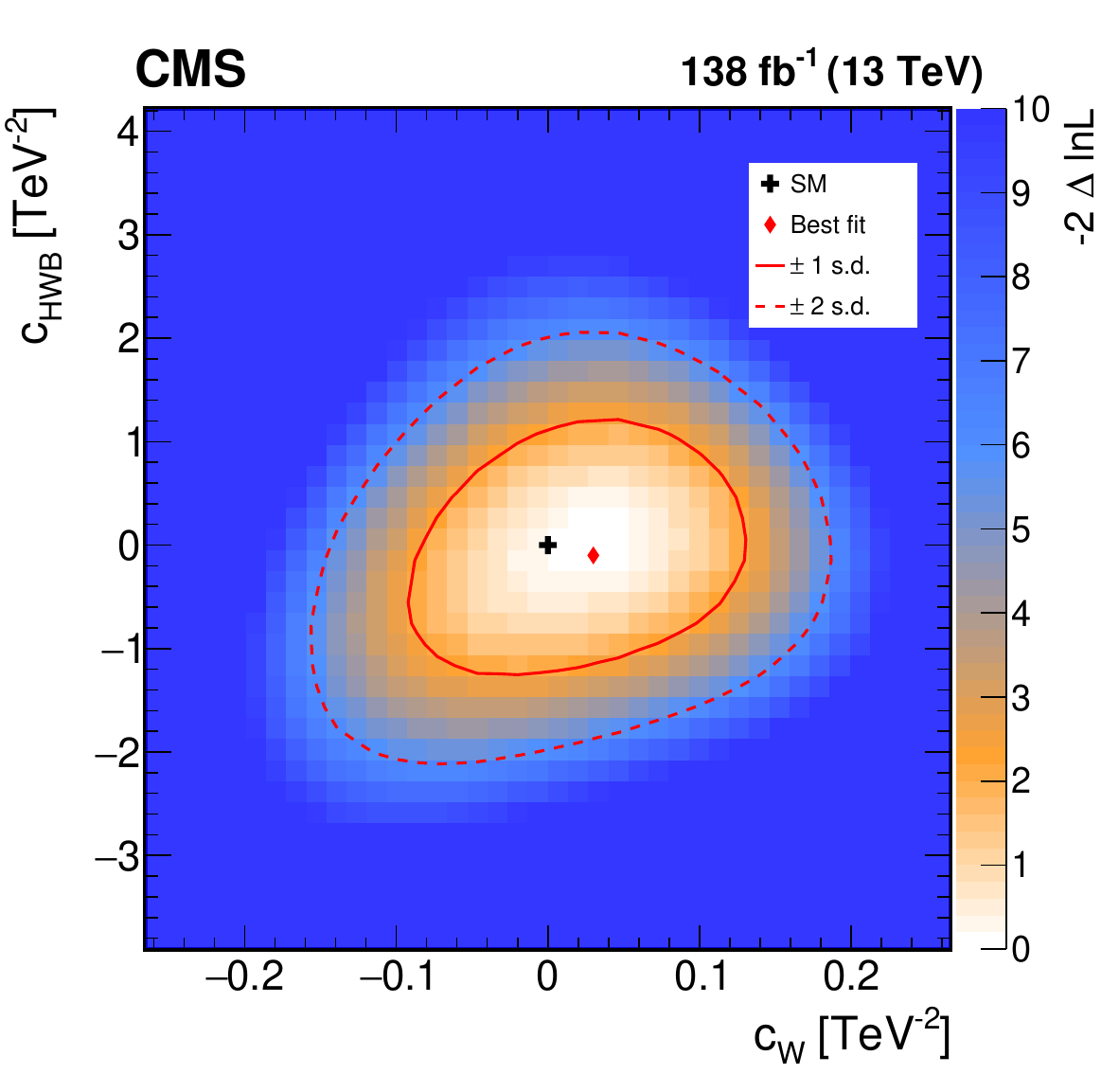}
      \caption{Negative of twice in the difference in the log-likelihood as a function of $\cw$ and $\chwb$ based on 138\fbinv of CMS data at 13\TeV. Upper left: the one-dimensional likelihood scan for $\cw$, showing the observed (black solid line) and expected (red dashed line) standard values, with 68\% and 95\% confidence intervals indicated by horizontal dashed lines. Upper right: the one-dimensional likelihood scan for $\chwb$, similarly presenting observed and expected limits. Lower: the two-dimensional likelihood contour for $\cw$ and $\chwb$, indicating the standard model (black cross), the best fit values (red dot), and contours corresponding to one (red solid line) and two (red dashed line) standard deviations.}
      \label{fig:EFT}
\end{figure*}

A maximum likelihood fit is performed using the DNN distribution with systematic uncertainties included in the fit as nuisance parameters. The distribution of the DNN output after the fit is shown in Fig.~\ref{fig:dnn}. The data agree with predictions though, in the lower panel, a slight trend is visible for the ratio of data to simulation in the last four bins, reaching the maximum value of 1.2 standard deviations in the highest DNN scores. The expected and observed 95\% confidence level intervals for \cw are, respectively, $[-0.21,0.25]$ and $[-0.11,0.16]$. For \chwb, the expected confidence level interval is $[-1.9,1.6]$ while it is observed to be $[-1.6,1.5]$. The one- and two-dimensional likelihood scans for \cw and \chwb are shown in Fig.~\ref{fig:EFT}. Results are in agreement with the SM expectation. Similar WCs can be studied in EW {\PZ}jj production as performed by ATLAS and CMS experiments~\cite{Aad2021_EFT,CMS:2019ppl}. The expected constraints on \cw in this analysis are found to be weaker than those in the CMS analysis of Ref.~\cite{CMS:2019ppl} while both \cw and \chwb are better constrained in comparison with ATLAS results.

\section{Summary}
\label{summary}
The first observation has been presented of the electroweak production of a photon in association with two jets (EW \ajj) using proton-proton collisions at $\sqrt{s}=13\TeV$ recorded with the CMS detector in 2016--2018 and corresponding to an integrated luminosity of 138\fbinv. Events are selected by requiring a photon with transverse momentum  $\ptg > 200 \GeV$ and two jets separated by at least $\abs{\Delta \eta} > 2.5$ with an invariant mass $\mjj > 500 \GeV$. The measured inclusive EW \ajj cross section is \xsec in agreement with the predicted standard model cross section of $177^{+13}_{-12} \unit{fb}$. Normalized differential cross sections are also measured as functions of several observables and compared with standard model predictions at next to leading order in perturbative quantum chromodynamics. Within the uncertainties, predictions agree with measurements in all observables except the pseudorapidity of the tagging jets. In particular, measured normalized cross sections differ from prediction by about two standard deviations in the pseudorapidity distribution of the softer tagging jet. The gap fraction is measured in a signal-enriched region and is found to be in agreement with the prediction, supporting the accuracy of the modeling of hadronic activities in VBF-like processes. A deep neural network is trained to probe new {\PW}{\PW}{\PGg} interactions in the context of an effective field theory, described by dimension-6 operators. The observed 95\% confidence intervals for the Warsaw basis Wilson coefficients $\cw$ and $\chwb$ are $[-0.11,0.16]$ and $[-1.6,1.5]$, respectively. 

\begin{acknowledgments}
We congratulate our colleagues in the CERN accelerator departments for the excellent performance of the LHC and thank the technical and administrative staffs at CERN and at other CMS institutes for their contributions to the success of the CMS effort. In addition, we gratefully acknowledge the computing centers and personnel of the Worldwide LHC Computing Grid and other centers for delivering so effectively the computing infrastructure essential to our analyses. Finally, we acknowledge the enduring support for the construction and operation of the LHC, the CMS detector, and the supporting computing infrastructure provided by the following funding agencies: SC (Armenia), BMBWF and FWF (Austria); FNRS and FWO (Belgium); CNPq, CAPES, FAPERJ, FAPERGS, and FAPESP (Brazil); MES and BNSF (Bulgaria); CERN; CAS, MoST, and NSFC (China); MINCIENCIAS (Colombia); MSES and CSF (Croatia); RIF (Cyprus); SENESCYT (Ecuador); ERC PRG, TARISTU24-TK10 and MoER TK202 (Estonia); Academy of Finland, MEC, and HIP (Finland); CEA and CNRS/IN2P3 (France); SRNSF (Georgia); BMFTR, DFG, and HGF (Germany); GSRI (Greece); NKFIH (Hungary); DAE and DST (India); IPM (Iran); SFI (Ireland); INFN (Italy); MSIT and NRF (Republic of Korea); MES (Latvia); LMTLT (Lithuania); MOE and UM (Malaysia); BUAP, CINVESTAV, CONACYT, LNS, SEP, and UASLP-FAI (Mexico); MOS (Montenegro); MBIE (New Zealand); PAEC (Pakistan); MES, NSC, and NAWA (Poland); FCT (Portugal); MESTD (Serbia); MICIU/AEI and PCTI (Spain); MOSTR (Sri Lanka); Swiss Funding Agencies (Switzerland); MST (Taipei); MHESI (Thailand); TUBITAK and TENMAK (T\"{u}rkiye); NASU (Ukraine); STFC (United Kingdom); DOE and NSF (USA).

\hyphenation{Rachada-pisek} Individuals have received support from the Marie-Curie program and the European Research Council and Horizon 2020 Grant, contract Nos.\ 675440, 724704, 752730, 758316, 765710, 824093, 101115353, 101002207, 101001205, and COST Action CA16108 (European Union); the Leventis Foundation; the Alfred P.\ Sloan Foundation; the Alexander von Humboldt Foundation; the Science Committee, project no. 22rl-037 (Armenia); the Fonds pour la Formation \`a la Recherche dans l'Industrie et dans l'Agriculture (FRIA) and Fonds voor Wetenschappelijk Onderzoek contract No. 1228724N (Belgium); the Beijing Municipal Science \& Technology Commission, No. Z191100007219010, the Fundamental Research Funds for the Central Universities, the Ministry of Science and Technology of China under Grant No. 2023YFA1605804, the Natural Science Foundation of China under Grant No. 12061141002, 12535004, and USTC Research Funds of the Double First-Class Initiative No.\ YD2030002017 (China); the Ministry of Education, Youth and Sports (MEYS) of the Czech Republic; the Shota Rustaveli National Science Foundation, grant FR-22-985 (Georgia); the Deutsche Forschungsgemeinschaft (DFG), among others, under Germany's Excellence Strategy -- EXC 2121 ``Quantum Universe" -- 390833306, and under project number 400140256 - GRK2497; the Hellenic Foundation for Research and Innovation (HFRI), Project Number 2288 (Greece); the Hungarian Academy of Sciences, the New National Excellence Program - \'UNKP, the NKFIH research grants K 131991, K 133046, K 138136, K 143460, K 143477, K 146913, K 146914, K 147048, 2020-2.2.1-ED-2021-00181, TKP2021-NKTA-64, and 2021-4.1.2-NEMZ\_KI-2024-00036 (Hungary); the Council of Science and Industrial Research, India; ICSC -- National Research Center for High Performance Computing, Big Data and Quantum Computing, FAIR -- Future Artificial Intelligence Research, and CUP I53D23001070006 (Mission 4 Component 1), funded by the NextGenerationEU program (Italy); the Latvian Council of Science; the Ministry of Education and Science, project no. 2022/WK/14, and the National Science Center, contracts Opus 2021/41/B/ST2/01369, 2021/43/B/ST2/01552, 2023/49/B/ST2/03273, and the NAWA contract BPN/PPO/2021/1/00011 (Poland); the Funda\c{c}\~ao para a Ci\^encia e a Tecnologia, grant CEECIND/01334/2018 (Portugal); the National Priorities Research Program by Qatar National Research Fund; MICIU/AEI/10.13039/501100011033, ERDF/EU, "European Union NextGenerationEU/PRTR", and Programa Severo Ochoa del Principado de Asturias (Spain); the Chulalongkorn Academic into Its 2nd Century Project Advancement Project, the National Science, Research and Innovation Fund program IND\_FF\_68\_369\_2300\_097, and the Program Management Unit for Human Resources \& Institutional Development, Research and Innovation, grant B39G680009 (Thailand); the Kavli Foundation; the Nvidia Corporation; the SuperMicro Corporation; the Welch Foundation, contract C-1845; and the Weston Havens Foundation (USA).
\end{acknowledgments}
\bibliography{auto_generated}

\cleardoublepage \appendix\section{The CMS Collaboration \label{app:collab}}\begin{sloppypar}\hyphenpenalty=5000\widowpenalty=500\clubpenalty=5000\input{SMP-19-005-public-authorlist.tex}\end{sloppypar}
\end{document}

%% file: SMP-19-005-public-authorlist.tex
\cmsinstitute{Yerevan Physics Institute, Yerevan, Armenia}
{\tolerance=6000
A.~Hayrapetyan, V.~Makarenko\cmsorcid{0000-0002-8406-8605}, A.~Tumasyan\cmsAuthorMark{1}\cmsorcid{0009-0000-0684-6742}
\par}
\cmsinstitute{Institut f\"{u}r Hochenergiephysik, Vienna, Austria}
{\tolerance=6000
W.~Adam\cmsorcid{0000-0001-9099-4341}, J.W.~Andrejkovic, L.~Benato\cmsorcid{0000-0001-5135-7489}, T.~Bergauer\cmsorcid{0000-0002-5786-0293}, M.~Dragicevic\cmsorcid{0000-0003-1967-6783}, C.~Giordano, P.S.~Hussain\cmsorcid{0000-0002-4825-5278}, M.~Jeitler\cmsAuthorMark{2}\cmsorcid{0000-0002-5141-9560}, N.~Krammer\cmsorcid{0000-0002-0548-0985}, A.~Li\cmsorcid{0000-0002-4547-116X}, D.~Liko\cmsorcid{0000-0002-3380-473X}, M.~Matthewman, I.~Mikulec\cmsorcid{0000-0003-0385-2746}, J.~Schieck\cmsAuthorMark{2}\cmsorcid{0000-0002-1058-8093}, R.~Sch\"{o}fbeck\cmsAuthorMark{2}\cmsorcid{0000-0002-2332-8784}, D.~Schwarz\cmsorcid{0000-0002-3821-7331}, M.~Shooshtari\cmsorcid{0009-0004-8882-4887}, M.~Sonawane\cmsorcid{0000-0003-0510-7010}, W.~Waltenberger\cmsorcid{0000-0002-6215-7228}, C.-E.~Wulz\cmsAuthorMark{2}\cmsorcid{0000-0001-9226-5812}
% ...
\par}
\vskip\cmsinstskip
\dag:~Deceased\\
$^{1}$Also at Yerevan State University, Yerevan, Armenia\\
$^{2}$Also at TU Wien, Vienna, Austria\\
% ...
